\documentclass{aastex631}

\usepackage{bm}
\usepackage{subcaption}
\usepackage[T1]{fontenc}
\usepackage{booktabs}
\usepackage{threeparttable}
\usepackage{multirow}
\usepackage{amsmath}
\usepackage[utf8]{inputenc}
\usepackage{makecell}
\usepackage{numprint}
\usepackage[ruled,vlined]{algorithm2e}
\usepackage{tabularx}

\begin{document}

 \title{A cellular automata approach for modelling pedestrian-vehicle mixed traffic flow in urban city}
 
\author{Jinghui Wang}

\affiliation{School of Safety Science and Emergency Management\\
Wuhan University of Technology\\
Wuhan, China}

\author{Wei Lv}
\altaffiliation{{\url{weil@whut.edu.cn} (W. Lv)}}
\affiliation{School of Safety Science and Emergency Management\\
Wuhan University of Technology\\
Wuhan, China}

\author{Yajuan Jiang}

\affiliation{School of Safety Science and Emergency Management\\
Wuhan University of Technology\\
Wuhan, China}

\author{Guangchen Huang}

\affiliation{School of Safety Science and Emergency Management\\
Wuhan University of Technology\\
Wuhan, China}

\begin{abstract}

In urban streets, the intrusion of pedestrians presents significant safety challenges. Modelling mixed pedestrian-vehicle traffic is complex due to the distinct motion characteristics and spatial dimensions of pedestrians and vehicles, making unified modelling difficult, with few studies addressing these issues. This paper employs a multi-grid cellular automata model to bridge the gap between vehicle and pedestrian models. An Improved Kerner-Klenov-Wolf (IKKW) model and a pedestrian motion model that incorporates Time-To-Collision (TTC) are introduced. Both models update the spatial motions of vehicles and pedestrians uniformly. Empirical analysis indicates that the model achieves high simulation accuracy. This model effectively illustrates the impact of pedestrian intrusion within mixed traffic scenario. The fundamental diagram of heterogeneous traffic reveals substantial differences, highlighting the effects of pedestrian intrusion on traffic flow states and identifying six phase regions in mixed traffic. Additionally, this paper examines conflicts between pedestrians and vehicles under varying speed limits and sidewalk widths, demonstrating that lower speeds and broader sidewalks significantly reduce the frequency of pedestrian-vehicle conflicts. Notably, the frequency of peak conflicts at a vehicle speed limit of 60.48 km/h is more than three times higher than at 30.24 km/h. This model offers a potential approach to studying mixed traffic flows and exhibits substantial scalability.

\end{abstract}


\section{Introduction} 
\label{section1}
The complex traffic environment is an important cause of frequent traffic accidents. More than 1.2 million people die in road crashes yearly, and more than 20 million are severely injured, making it the 9th leading cause of death globally (2.2 \% of all deaths globally) \citep{world2015global}. Pedestrian deaths comprise more than 35\% of road accident deaths, mainly due to pedestrian-vehicle crashes \citep{hacohen2018dynamic}. In cities, mixed traffic scenarios are common, and there are significant differences in driving speed and size between motor vehicles, non-motor vehicles, and pedestrians. Compared with the homogeneous flow, the traffic conflicts of mixed traffic flow will appear more frequently, causing more frequency of accidents. The irregular driving behaviors and complex traffic environments are important reasons for traffic accidents.

In the research of pedestrian-vehicle mixed traffic flow, many scholars have analyzed the videos, questionnaire surveys, and empirical data to study the microscopic movement characteristics of pedestrians and vehicles, the distribution of conflicting locations, etc. \citep{brewer2006exploration,yang2006modeling,almodfer2016quantitative,alhajyaseen2017studying,ferenchak2018spontaneous,jiang2019drivers,chen2019assessing,kalatian2021decoding,bustos2021explainable}. \citet{brewer2006exploration} investigated 42 observation points in seven states in the United States and found that pedestrians did not always wait to cross the street when all lanes were completely clear. Instead, they anticipated that the lanes would clear as they cross and used a "rolling gap" to cross the street. The 11 approaches (pedestrian’s approach to a crossing) had 85th percentile accepted gaps between 5.3 and 9.4 s, with a trend of increasing gap length as crossing distance increased. \citep{yang2006modeling} established a model of pedestrians crossing the road, using questionnaire surveys and video data analysis methods to verify the model's parameters, and research the gap acceptance and arrival distribution of different pedestrian types (law-obeying ones and opportunistic ones). \citet{almodfer2016quantitative} analyzed the video data and found that the far lanes recorded more serious conflicts than the near lanes. Besides, they observed that when the waiting time for pedestrians decreases, the frequency of serious conflicts will also decrease. \citet{alhajyaseen2017studying} observed the collision of people and vehicles at intersections and found that the location distribution of acceleration events is concentrated at the entrance points to the pedestrian-vehicle conflict area for both far-side and near-side pedestrians.  \citet{ferenchak2018spontaneous} 's research showed that the spontaneous order of space begins shifting to a new leader at approximately a 1:1 ratio of the different modes. This finding means that as pedestrians outnumber vehicles in space, the vehicles will begin to yield to the pedestrians. As the vehicles reach higher numbers than the pedestrians, the pedestrians will again give way to the vehicles. \citet{jiang2019drivers} found that the size of the deceleration zone in China is smaller than that in Germany and recommended 2 m as the minimum distance threshold for collision recognition. In their research, \citet{chen2019assessing} pointed out that the larger the intersection size and corner, the worse the safety performance. \citet{kalatian2021decoding} used virtual reality (VR) to conduct pedestrian crossing experiments and investigated various factors of pedestrians' wait time before crossing unsignalized crosswalks. \citet{bustos2021explainable} used computer vision (CV) techniques to identify the risks to people and vehicles in the street scene and establish a fine-grained map of hazard levels across a city.
In addition to the above research, many models applied to pedestrian-vehicle interaction have emerged. \citet{zhang2014dynamic} established a cell transmission model to model pedestrian-vehicle evacuation during large-scale evacuation and used optimal signal timings at the right turn of the road to reduce congestion and queuing. \citet{karaaslan2018modeling} used AnyLogic to build a microscopic pedestrian-vehicle simulation model. It is pointed out that compared to internal combustion engine vehicles (ICEVs), electric vehicles (EVs) have more significant vehicle-pedestrian traffic safety risks due to their lower noise, and the magnitude of the risk is related to ambient sound levels and illumination. \citet{hacohen2018dynamic} presented a novel model for pedestrian path planning when sharing the road with vehicles. According to the model, pedestrians examine the traffic in the vicinity of the crossing zone before crossing and construct a virtual probability map for trajectory planning. These models include macroscopic cell transmission models, microscopic social force models, and agent-based models, which can simulate and analyze the conflicts and interactions between pedestrians and vehicles.

An abstraction of traffic elements in a discrete way provides a unique advantage for using cellular automata approach to carry out traffic research. Cellular automata have been widely used to study traffic flow in recent decades. In the research of mixed traffic flow using the cellular automata methods, the following scholars mainly focus on heterogeneous motor vehicles \citep{kong2021modeling,meng2011improved,ruan2017improved,jia2005effect,yang2015cellular,mu2012analysis,dailisan2020crossover}, motor vehicles and non-motor vehicles \citep{dong2020traffic,vasic2012cellular,ren2016heterogeneous,lan2010cellular,luo2015modeling,radhakrishnan2013hybrid}, heterogeneous non-motor vehicles \citep{jiang2014properties,jia2007multi,jin2015improved}, and motor vehicles and pedestrians \citep{jiang2006interaction,xie2012cellular,xin2014power,li2015studies,zhao2016cellular,echab2016simulation,layegh2020modeling,li2021safety,lu2016cellular}. Tab.\ref{table1} provides an overview of a selection of relevant studies.

\begin{table}[ht]
\centering
\footnotesize
\caption{Overview of mixed traffic flow modelling by cellular automata.}
\begin{tabular}{
  >{\centering\arraybackslash}m{2.5cm}
  >{\centering\arraybackslash}m{2.5cm}
  >{\centering\arraybackslash}m{3.5cm}
  >{\centering\arraybackslash}m{1.5cm}
  >{\centering\arraybackslash}m{3.5cm}
  >{\centering\arraybackslash}m{2cm}
}
\toprule
Author & Classes & Vehicle types & Cell length (m) & Simulation scenarios & Considering the conflict \\
\midrule
\citet{kong2021modeling} & \multirow{7}{=}{Heterogeneous motor vehicles} & Car and truck & 0.5 & Two-lane road & No \\
\citet{meng2011improved} &  & Car and truck & 0.5 & Three-lane road considering heterogeneous work zone & No \\
\citet{ruan2017improved} &  & Vehicles with different lengths and weight & 5 & Two-lane long-span bridge & No \\
\citet{jia2005effect} &  & Vehicles with different lengths and speed & 7.5 & Single-lane road & No \\
\citet{yang2015cellular} &  & Car and truck & 0.5 & Single-lane road & No \\
\citet{mu2012analysis} &  & Conventional car and micro-car & 4 & Two-lane road & No \\
\citet{dailisan2020crossover} &  & Bus and car & 7.5 & Two-lane road & No \\
\midrule
\citet{dong2020traffic} &  \multirow{6}{=}{Motor vehicles and non-motor vehicles} & Bicycle, electric bicycle, and minor vehicle & 2.5 & Mixed traffic three-lane road & Yes \\
\citet{vasic2012cellular} &  & Car and bicycle & 3.75, 7.5 & A T-shaped intersection & No \\
\citet{ren2016heterogeneous} &  & Car and bicycle & 1 & Mixed traffic signal intersection & No \\
\citet{lan2010cellular} &  & Car and motorcycle & 1 & Mixed traffic two-lane road & No \\
\citet{luo2015modeling} &  & Car and bicycle & 1 & Mixed traffic two-lane road & No \\
\citet{radhakrishnan2013hybrid} &  & Rickshaw, bicycle, car, bus, etc. & 1 & A section of road with no lane division & No \\
\midrule
\citet{jiang2014properties} &  \multirow{3}{=}{Heterogeneous non-motor vehicles} & Bicycle and electric bicycle & 1 & Non-motor three-lane road & No \\
\citet{jia2007multi} &  & Fast and slow bicycle & 4 & Single-lane nonmotor road & No \\
\citet{jin2015improved} &  & Bicycle and electric bicycle & 2 & Single-lane nonmotor road & No \\
\midrule
\citet{jiang2006interaction} &  \multirow{7}{=}{Motor vehicles and pedestrians} & Vehicle and pedestrian & 0.4 & Channel & No \\
\citet{xie2012cellular} &  & Vehicle and pedestrian & 3.75 & Signalized Crosswalk & No \\
\citet{xin2014power} &  & Vehicle and pedestrian & 0.4 & Midblock crosswalk & No \\
\citet{li2015studies} &  & Vehicle and pedestrian & 0.6 & Two-lane road & No \\
\citet{zhao2016cellular} &  & Large vehicles, Passenger cars, and pedestrian & 0.6 & Non signalized mid-block crosswalks & No \\
\citet{echab2016simulation} &  & Vehicle and pedestrian & 0.6 & A single-lane roundabout & No \\
\citet{layegh2020modeling} &  & Vehicle and pedestrian & 0.6 & A multi-lane roundabout & Yes \\
\citet{li2021safety} &  & Vehicle and pedestrian & 1.2, 3.6 & Non signalized mid-block crosswalks & Yes \\
\citet{lu2016cellular} &  & Vehicle and pedestrian & 0.4, 7.5 & Non signalized mid-block crosswalks & Yes \\
\bottomrule
\end{tabular}
\label{table1}
\end{table}

In the study of pedestrian-vehicle mixed traffic, \citet{jiang2006interaction} used the lattice gas model to study the pedestrian-vehicle interaction in narrow passages. Research points out that for a given number of pedestrians, there is a critical maximum speed, below which the vehicle will be affected by pedestrians. The critical maximum speed increases with the increase in the number of pedestrians. \citet{xie2012cellular} studied the influence of different pedestrian types, signal cycle time, and green signal ratio on the flow of pedestrians and motor vehicles. \citet{xin2014power} established a pedestrian-vehicle mixed traffic model, considering two driving modes of the vehicles. The impact of different numbers of pedestrians crossing the road on vehicles and the changes in the cluster size of different types of pedestrians crossing the road are studied. It is found that when the system is composed of homogeneous pedestrians at the same time, the traffic flow tends to be "polarized" (the cluster sizes of pedestrians' group crossing either distributed in a wide area or only in small sizes), and when the system contains heterogeneous pedestrians, the traffic flow tends to be "centralized" (the cluster size obeys a power-law distribution). \citet{li2015studies} studied the lane-changing behavior caused by vehicles avoiding pedestrians. Simulation results show that lane-changing behavior can improve traffic efficiency and reduce pedestrian-vehicle conflicts. The negative effect of lane-changing is that pedestrians must stay longer between the lanes in the crossing. \citet{zhao2016cellular} found that the pedestrian crossing has less impact on traffic when the vehicles’ density is lower than 25 veh/km/lane. When the vehicles’ density is higher than 60 veh/km/lane, pedestrians' crossing behavior at the crosswalk without signal control will significantly affect the traffic flow. \citet{echab2016simulation} introduced Nagel-Schreckenberg (NaSch) rules \citep{nagel1992cellular} to simulate one-way lane changes and crosswalks at the entrance of roundabouts. The impact of the location of the crosswalk on traffic is studied. \citet{layegh2020modeling} consider different types of pedestrian crossing behaviors, pointing out middle-aged/elderly pedestrians, compared to the teenager/young people, are 7\% less prone to conflict with vehicles when crossing the roundabout. In addition, the likelihood of conflicts is reduced by 26\% and 8\% when the pedestrians and vehicles had lower speeds, respectively. \citet{li2021safety} used fuzzy cellular automata to simulate the conflicting relationship between pedestrians and vehicles. The study found that a higher probability of a driver giving way improves the safety of pedestrians crossing the zebra crossing and reduces traffic efficiency. The increase in traffic flow hurts pedestrian safety and traffic efficiency.

The above literature review shows that studying pedestrian-vehicle mixed traffic is a hot topic in traffic research. However, most existing research focuses on pedestrian-vehicle conflicts at crosswalks or right-turn lanes at intersections, and there are fewer studies on conflicts in mixed traffic. The motor vehicle lanes are generally narrow in campus areas or city streets, and sidewalks are generally set up beside the motor vehicle lanes. In the case of high pedestrians’ density, pedestrians may invade the motor lane, and there is little research on mixed traffic in this scenario. To this end, this paper models pedestrian-vehicle mixed traffic via the cellular automata model and researches traffic conflict and congestion in this scenario.

The remainder of this paper is structured as follows. Section~\ref{section2} introduces the research scenario and the modeling concepts discussed in this study. Section~\ref{section3} details the proposed model and provides preliminary validation and analysis. Section~\ref{section4} describes the simulation analyses and discusses scenarios involving mixed pedestrian and vehicle traffic. Section~\ref{section5} presents the fundamental and phase diagrams for mixed traffic flow. Section~\ref{section6} explores the variation in conflicts under different speed limits and sidewalk widths. The paper concludes with Section~\ref{section7}, which summarizes the findings.

\section{Scenario} \label{section2}
\begin{figure}[ht!]
\centering
\includegraphics[scale=0.5]{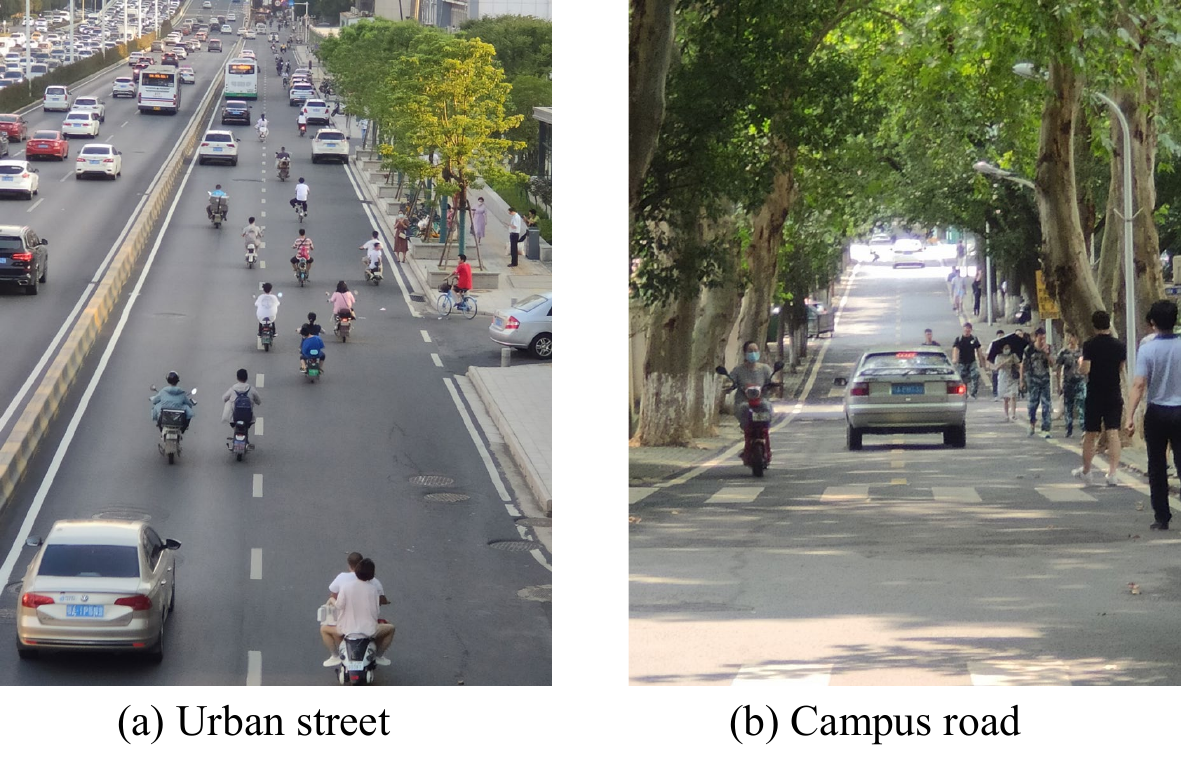}
\caption{Mixed traffic scenes in urban streets and campus roads, Captured on July, 2021, Location: Luoshi Road (a) and Wuhan university of technology (b), Wuhan.}
\label{fig1}
\end{figure}

Due to space constraints in many city streets, it is common for pedestrians or non-motor vehicles to invade the motorway. This behavior significantly reduces the traffic efficiency of motor vehicles. Moreover, in mixed traffic, compared with homogeneous traffic, there will be more conflicts between different vehicles and pedestrians, which can easily cause traffic accidents. Fig.\ref{fig1} shows a typical mixed traffic scene on city streets and campus roads.

\begin{figure}[ht!]
\centering
\includegraphics[scale=0.3]{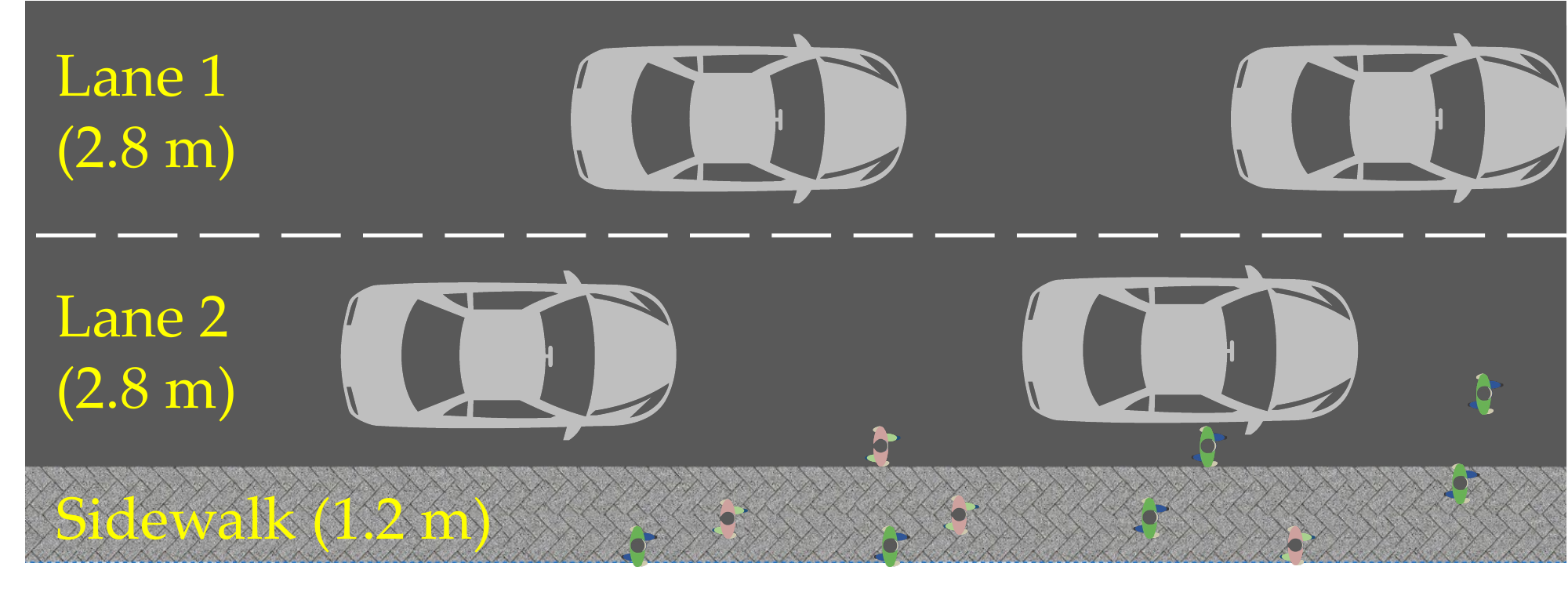}
\caption{Schematic diagram of pedestrian and vehicle mixed traffic scene.}
\label{fig2}
\end{figure}

Our research focus on the scene of pedestrian intrusion into the motorway, as shown in Fig.\ref{fig2}. Two-lane roads in the same direction are adjacent to the sidewalk. Due to the excessively high pedestrians’ density and narrow sidewalk design, pedestrians are prone to invade the motorway to accelerate or avoid other pedestrians. Therefore, the vehicles have to slow down or change lanes to avoid them.

%%%%%fig2

\begin{figure}[ht!]
\centering
\includegraphics[scale=0.4]{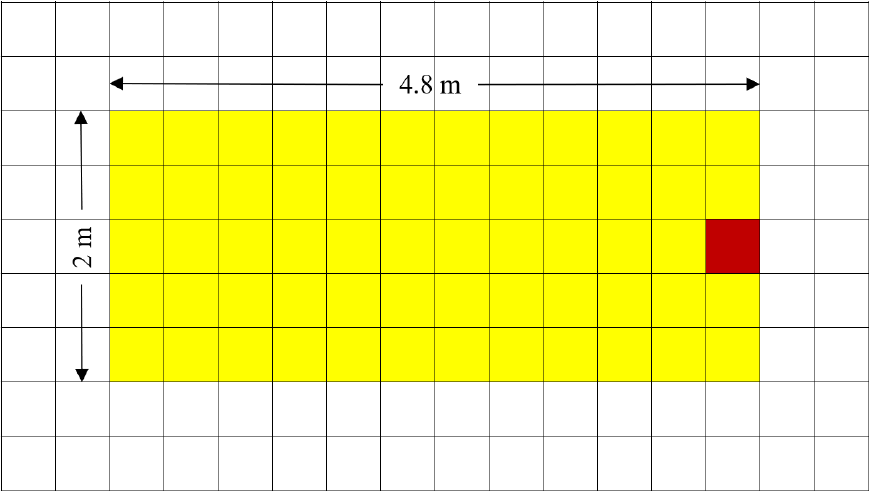}
\caption{Vehicle representation using a multi-grid model (red indicates the driver cell, yellow indicates the carrier cells).}
\label{fig3}
\end{figure}

\begin{figure}[ht!]
\centering
\includegraphics[scale=0.45]{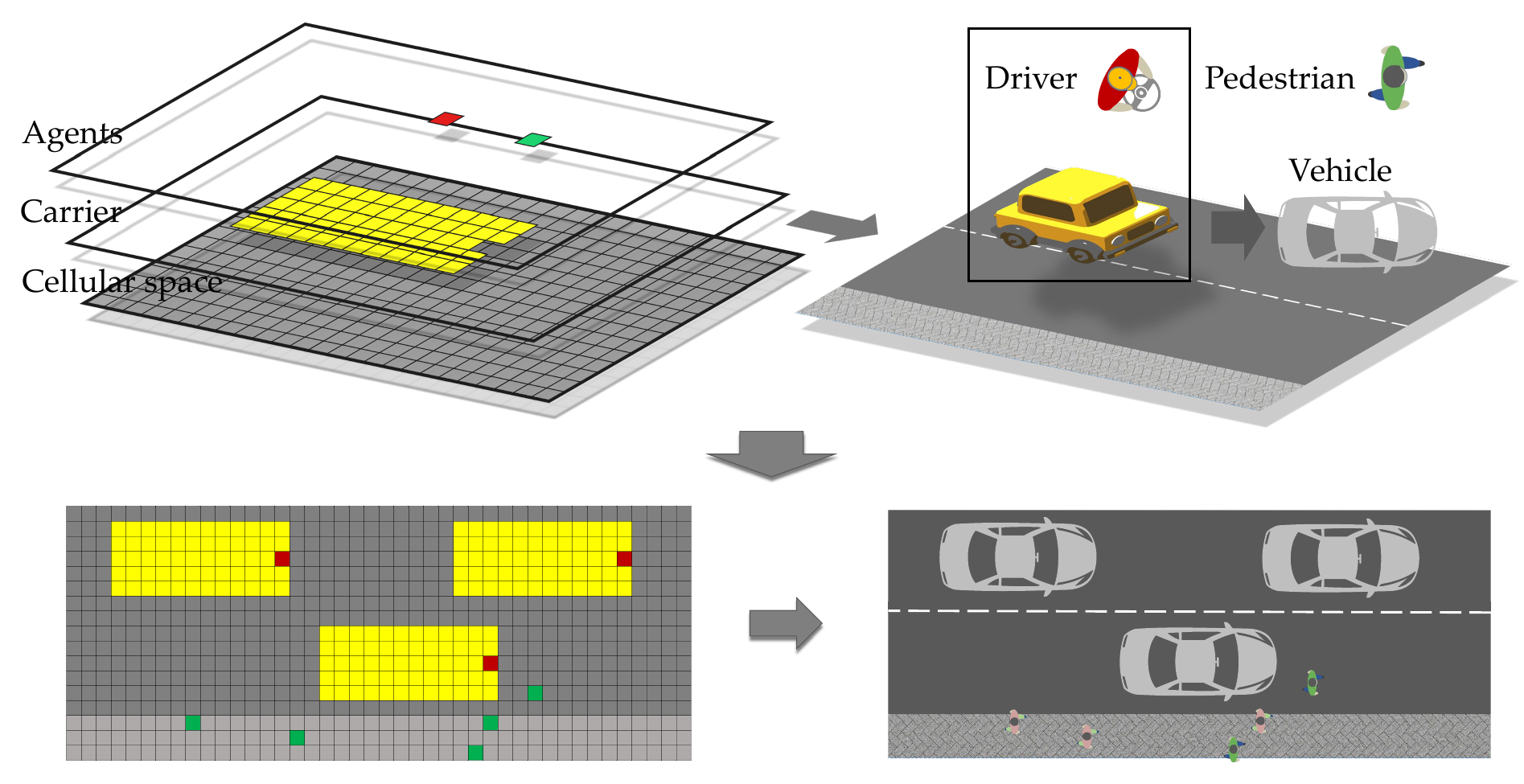}
\caption{Vehicle-pedestrian mixed traffic model (the green cell represents a single pedestrian).}
\label{fig4}
\end{figure}

This paper attempts to simulate and analyze the pedestrian-vehicle mixed traffic of the scenario in Fig.\ref{fig2} through the cellular automata model. The modeled road and sidewalk are represented by a grid of cells, each cell representing 0.4 m\(\cdot\)0.4 m square of the physical space, the width of the motor vehicle lane is 2.8 m (7 cells in length), and the sidewalk is 1.2 m (3 cells in length). Each pedestrian occupies one cell of space (0.4 m\(\cdot\)0.4 m), and each vehicle occupies 5\(\cdot\)12 cells of space (2 m\(\cdot\)4.8 m). Similar to the multi-grid model \citep{song2006simulation}, each vehicle in this paper is decomposed into multi-grid cells of space, consisting of one single driver cell and carrier cells (representing a vehicle without a driver). The schematic diagram is shown in Fig.\ref{fig3} and Fig.\ref{fig4}.

%%%%%%

\begin{figure}[ht!]
\centering
\includegraphics[scale=0.5]{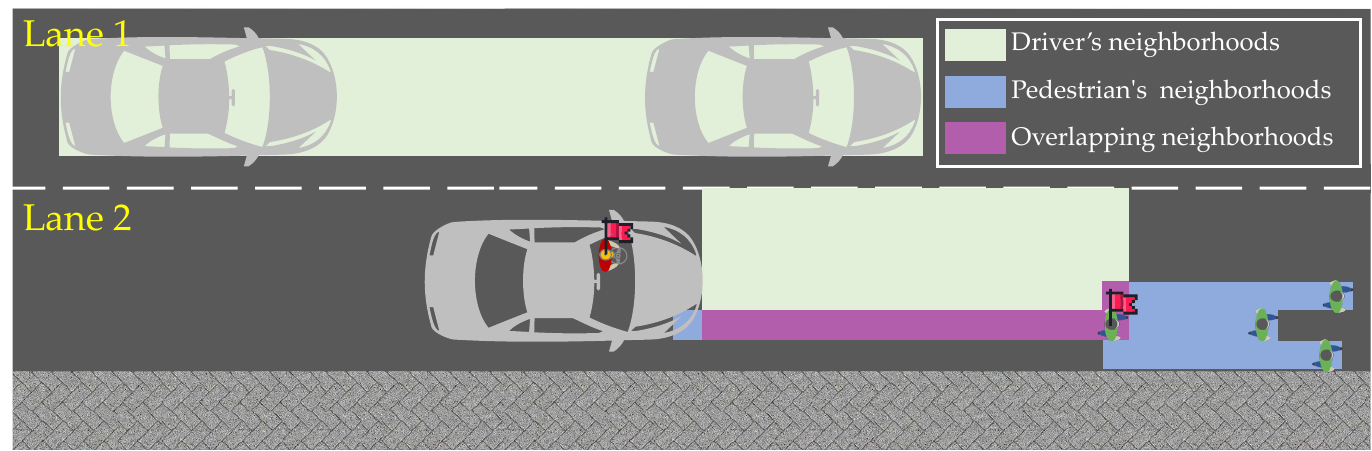}
\caption{Neighborhoods diagram of selected driver and pedestrian (red flags mark selected driver and pedestrians).}
\label{fig5}
\end{figure}

The model physical space is a cellular space consisting of \(m\cdot n\) (corresponding length and width) cells. The cells are defined as empty cells, driver cells, carrier cells, pedestrian cells according to their properties, and obey the principle of volume exclusion. Within the model update process, first, each cell is traversed in space through a loop statement to find agents (driver cells and pedestrian cells), and second, search their neighborhoods. The schematic diagram of the neighborhoods of drivers and pedestrians is shown in Fig.\ref{fig5}. Then, the drivers and pedestrians update their state according to the neighborhood information and   model rules in the next time timestep, uniformly. Lastly, the state update of the cellular space is performed (update the single driver cell and carrier cells as a whole (vehicle)). The modelling idea of the mixed traffic is introduced above, and in the following model declaration, the term "vehicle" (a collection of drivers and vehicles) is uniformly used.

\section{Model} \label{section3}

The pedestrian-vehicle mixed traffic model proposed in this paper comprises a vehicle movement sub-model and a pedestrian walking sub-model. A pedestrian-vehicle mixed traffic simulation model is established by coupling these two sub-models.

\subsection{Vehicle movement model} \label{subsection3.1}

\subsubsection{Car-following model} \label{subsection3.1.1}

Among the traffic flow models established by the cellular automata method, the most widely used is the NaSch rule proposed by Nagel and Schreckenberg \citep{nagel1992cellular}. This model is simple and easy to modify, can simulate traffic jams, stop-and-go waves, and reproduce the fundamental diagram characteristics via simple rules. However, due to the coarse cell division in the model (a cell length is 7.5 m), the model can hardly simulate the microscopic position and speed changes of the vehicle. After proposing the three-phase traffic flow theory, \citet{kerner2002cellular} employed the Kerner-Klenov-Wolf (KKW) model to validate it. In this model, the cell length represents 0.5 m, and with this representation, the model can reproduce various synchronized flow phenomena. In addition, due to the finer-grid division of the KKW model, it is conducive to combining with the pedestrian model. Therefore, this paper improves the KKW model to establish the vehicle's car-following rules.

In the KKW model, four versions (KKW-1~KKW-4) are proposed according to different model parameters and variations, and this paper presents an improvement of the KKW-1 model. In the KKW-1 model, constant acceleration and synchronization distance are set to finely describe the continuous change of vehicle speed in synchronized flow, which causes the long acceleration period and forced braking rules of vehicles in this model to be inconsistent with reality. Therefore, improvements to the KKW-1 model are necessary for modelling mixed traffic flows with strong velocity oscillation phenomena. This paper proposes an Improved Kerner-Klenov-Wolf (IKKW) model and makes the following variations:

\textbf{(a) Dynamic Acceleration:} Unlike the constant acceleration setting in the KKW-1 model, in this model, the vehicle's acceleration fluctuates within the range \(({{a}_{\min }}\sim{{a}_{\max }})\) according to its speed difference from the leading vehicle (whether it is another vehicle or a pedestrian). Therefore, the vehicle's motion characteristics during the approach to and departure from congestion areas are consistent with actual car-following situations.

\textbf{(b) Dynamic Synchronization Distance:} The constant synchronization distance setting in the KKW-1 model leads to a phenomenon where high-speed vehicles cannot decelerate in time when approaching congestion areas, which contradicts the actual car-following motion process. In this model, a dynamic synchronization distance is set. In synchronized flow, vehicles maintain a small synchronization distance, whereas when approaching congestion areas, the synchronization distance increases accordingly.

Furthermore, in order to ensure compatibility with pedestrian cells, we have adjusted the representation of cell length from 0.5 m to 0.4 m. The speed update process for the target vehicle is categorized into two parts: dynamical and stochastical speed updates. The dynamical part involves deterministic speed adjustments based on the speed and distance of the leading vehicle (whether it is a vehicle or a pedestrian). On the other hand, the stochastical part accounts for the random acceleration or deceleration process (noise) of the target vehicle and represents a non-deterministic speed update. The rules of the IKKW model are outlined in Eqs.\ref{1}-\ref{10}.

\textbf{1. Dynamical part:}

• Determine the acceleration \(a\) of the target vehicle.

\begin{equation}\label{1}
a = \left\{ \begin{array}{l}
{a_{\min }}{\rm{                               \qquad\qquad\qquad\qquad\quad\quad\;\:\; if\:}}\left|(v_{tar}^t - v_l^t)\right| < \lceil({a_{\max }}{\rm{/2)}}\rceil\\
\lceil({a_{\max }} \cdot (v_{tar}^t - v_l^t)/{v_{free}}{\rm{)\rceil   \quad   if\:}}(v_{tar}^t - v_l^t) > \lceil({v_{free}}{\rm{/2)  }}\rceil\\
\lceil({a_{\max }}{\rm{/2)\rceil \qquad \qquad\quad\quad\quad\quad  otherwise }}
\end{array} \right.
\end{equation}

Eq.\ref{1} describes the process of determining the vehicle acceleration, where \(v_{tar}^{t}\) represents the speed of the target vehicle at the \(t\)-th timestep, and \(v_{l}^{t}\) represents the speed of the leader one (vehicle or pedestrian) and \({{v}_{free}}\) represents the free flow velocity (maximum velocity), which is a constant.  (\({{a}_{\min }}=\delta a\)), the acceleration unit \(\delta a\), the smallest speed unit \(\delta v\), the smallest distance unit \(\delta x\) and the timestep unit (\(\tau \)) satisfy the following relation: \(\delta a=\delta v/\tau ,\text{ }\delta v=\delta x/\tau \)). \({{a}_{\min }}\) and \({{a}_{\max }}\) represent the minimum and maximum acceleration of the target vehicle, respectively.

The first term in Eq.\ref{1} states that when the speed difference between the target vehicle and the leader one (vehicle or pedestrian) is less than a certain threshold (here, we set it as \(\lceil({{a}_{\max }}\text{/2)}\rceil\)), then the acceleration of the vehicle at that moment is set as \({{a}_{\min }}\). This term allows the vehicle to maintain a small acceleration change when moving in synchronized flow to obtain a stable movement state. The second term in Eq.\ref{1} describes that when the speed of the target vehicle is higher than the leader one (vehicle or pedestrian) and exceeds a certain threshold (here, we set it as \(\lceil({{v}_{free}}\text{/2)}\rceil\)), the target vehicle will set its acceleration (\({{a}_{\min }}\tilde{\ }{{a}_{\max }}\)) according to its speed difference (the target vehicle and the leader one), corresponding to the early deceleration process of the vehicle as it approaches the congestion area. In this process, the value of its acceleration changes dynamically with the speed difference between the leader one. The third term in Eq.\ref{9} represents the constant acceleration (\(\lceil({{a}_{\max }}\text{/2)}\rceil\)) maintained under normal conditions.

• Calculate safe driving speed of the target vehicle at \(t\)-th timestep (\(v_{s}^{t}\)).

\begin{equation}\label{2}
v_s^t = {g_{c,front}}/\tau 
\end{equation}

In Eq.\ref{2}, \(v_{s}^{t}\) represents the safe driving speed of the target vehicle at \(t\)-th timestep. \({{g}_{c,front}}\) represents the bumper-to-bumper space gap between the target vehicle and its leader one in the current lane.

• Calculate the adaptation speed of the target vehicle at \(t\)-th timestep (\(v_{a}^{t}\)).

\begin{equation}\label{3}
v_a^t = \left\{ \begin{array}{l}
v_{tar}^t + a \cdot \tau {\rm{\qquad\qquad\quad\qquad\;for\:  }}{{\rm{g}}_{c,front}} > {k_1} \cdot v_{tar}^t \cdot \tau {\rm{ }}\\
v_{tar}^t + a \cdot \tau {\rm{\qquad\qquad\quad\qquad\;for\:  }}{k_2} \cdot v_{tar}^t \cdot \tau  < {{\rm{g}}_{c,front}} \le {k_1} \cdot v_{tar}^t \cdot \tau {\rm{\: and\: }}v_{tar}^t - v_l^t < \lceil(\lambda  \cdot {v_{free}})\rceil\\
v_{tar}^t{\rm{\qquad\qquad\qquad\qquad\qquad\;for\:  }}{{\rm{g}}_{c,front}} \le {k_2} \cdot v_{tar}^t \cdot \tau {\rm{\: and\: }}v_l^t < v_{tar}^t \le {a_{\max }}\\
v_{tar}^t + a \cdot \tau  \cdot {\mathop{\rm sgn}} (v_l^t - v_{tar}^t){\rm{\;\;otherwise }}
\end{array} \right.
\end{equation}

Different from the KKW-1 model, this model sets a different synchronization distance when calculating the adaptation speed, where \({{k}_{1}}\) and \({{k}_{2}}\) are dimensionless parameters in the model (\({{k}_{1}}>{{k}_{2}}\)), \({{k}_{1}}\cdot v_{tar}^{t}\cdot \tau \) represents the upper bound of the synchronization distance when car-following, \({{k}_{2}}\cdot v_{tar}^{t}\cdot \tau \) represents the upper bound of the synchronization distance in the synchronized flow. Three intervals appear to correspond to different bumper-to-bumper space gaps, corresponding to Eq.\ref{3}.

The first term of Eq.\ref{3} indicates that when the bumper-to-bumper space gap falls to the first interval (\({{\text{g}}_{c,front}}>{{k}_{1}}\cdot v_{tar}^{t}\cdot \tau \)), the vehicle accelerates without being affected by the leader one (vehicle or pedestrian). The second term of Eq.\ref{3} indicates that when the bumper-to-bumper space gap is located in the second interval (\({{k}_{2}}\cdot v_{tar}^{t}\cdot \tau <{{\text{g}}_{c,front}}\le {{k}_{1}}\cdot v_{tar}^{t}\cdot \tau \)) and satisfies the speed relationship (\(v_{tar}^{t}-v_{l}^{t}<ceil(\lambda \cdot {{v}_{free}})\)), the vehicle continues to accelerate. This term corresponds to the behavior of the target vehicle approaching a congestion area. If the speed difference between the target vehicle and the leader one (vehicle or pedestrian) is in a reasonable interval \(v_{tar}^{t}-v_{l}^{t}<\lceil(\lambda \cdot {{v}_{free}})\rceil\), the slowing down behavior is not necessary. The third term of Eq.\ref{3} indicates that when the bumper-to-bumper space gap is located in the third interval (\({{\text{g}}_{c,front}}\le {{k}_{2}}\cdot v_{tar}^{t}\cdot \tau \)), and the speed relationship (\(v_{l}^{t}<v_{tar}^{t}\le {{a}_{\max }}\)) is satisfied, the vehicle will maintain the original speed. This avoids premature deceleration behavior of vehicles approaching congestion areas, resulting in dense congestion areas. The fourth term in Eq.\ref{3} represents all cases where speed-adaptive behavior should be adopted, where the step function sgn(\(x\)) is 1 for \(x > 0\), 0 for \(x = 0\), and -1 for \(x < 0\).

• Calculate the speed of the dynamic part of the target vehicle at time \(t+1\) (\(\tilde{v}_{tar}^{t+1}\)).

\begin{equation}\label{4}
\tilde v_{tar}^{t + 1} = \max (0,\min ({v_{free}},v_s^t,v_a^t))
\end{equation}

Here, \(\max (x,y)\) is a function that takes the maximum value of \(x\) and \(y\).

\textbf{2. Stochastical part:}

The stochastical part of the model simulates the noise in the car-following.

• Determine the probability of target vehicle’s acceleration (\({{p}_{a}}\)).

\begin{equation}\label{5}
{p_a} = \left\{ \begin{array}{l}
{p_{a1}}\quad\text{if } \tilde{v}_{tar}^{t + 1} < {v_p}\\
{p_{a2}}\quad\text{if } \tilde{v}_{tar}^{t + 1} \ge {v_p}
\end{array} \right.
\end{equation}

\({{p}_{a1}}\) and \({{p}_{a2}}\) are dimensionless constants less or equal to 1, satisfying \({{p}_{a1}}>{{p}_{a2}}\). Eq.\ref{5} simulates the effect that the target vehicle moving at low speed in the dense flow tends to close up to the leading one. If the probability \({{p}_{a}}\) of the acceleration is high, the speed coordination effect is weak, with the probability \({{p}_{a}}\) the vehicle does not reduce its speed, and it may do so until it reaches the minimal safe gap. Note that the tendency to minimize the space gap at low speed can lead in particular to the "pinch" effect in synchronized traffic flow \citep{kerner1998experimental}, i.e., the self-compression of the synchronized flow at lower vehicles’ speed with the spontaneous emergence of moving jams. Therefore, \({{v}_{p}}\) can be regarded as the minimum speed at which the vehicles maintain a stable synchronized flow.

• Determine the probability of target vehicle’s deceleration (\({{p}_{b}}\)).

\begin{equation}\label{6}
{p_b} = \left\{ \begin{array}{l}
{p_0}{\rm{\quad \quad  if\:  }}v_{tar}^t = 0\\
1 - {p_a}{\rm{\;if \: }}\tilde v_{tar}^{t + 1} - v_l^t{\rm{ > }}{a_{\max }}\\
{p_1}{\rm{\quad\quad  if \:}}v_l^t < \tilde v_{tar}^{t + 1} \le {a_{\max }}{\rm{ }}\\
{p_2}{\rm{\quad\quad otherwise}}
\end{array} \right.
\end{equation}

In Eq.\ref{6}, \({{p}_{0}}\), \({{p}_{1}}\) and \({{p}_{2}}\) are dimensionless constants less or equal to 1, the first term in the equation is the slow-to-start rule for target vehicles, which occurs in almost all cellular automata car-following models. The second term in Eq.\ref{6} indicates that when the speed of the target vehicle and the speed of the leader one (vehicle or pedestrian) satisfy the relationship (\(\tilde{v}_{tar}^{t+1}-v_{l}^{t}\text{}{{a}_{\max }}\)), the target vehicle will decelerate with a greater probability. The third term of the equation serves the same purpose as the third term in Eq.\ref{3}; the fourth term represents the probability of deceleration in the normal state.

• Calculate the target vehicle's speed at time \(t+1\) (\(v_{tar}^{t\text{+}1}\)).

\begin{equation}\label{7}
\eta  = \left\{ \begin{array}{l}
 - 1{\rm{ \quad \;if\:  }}rand < {p_b}\\
1{\rm{  \qquad if\:  }}{p_b} \le rand < {p_b} + {p_a}\\
0{\rm{ \qquad  otherwise}}
\end{array} \right.
\end{equation}

\begin{equation}\label{8}
a'{\rm{ = }}\left\{ \begin{array}{l}
a{\rm{    \qquad if\:  }}v_{tar}^t{\rm{ = 0}}\\
{a_{\min }}{\rm{ \;\; if \: }}v_{tar}^t \ne {\rm{0}}
\end{array} \right.
\end{equation}

\begin{equation}\label{9}
v_{tar}^{t{\rm{ + }}1} = \max (0,\min (\tilde v_{tar}^{t + 1} + a' \cdot \tau  \cdot \eta ,v_{tar}^t + a \cdot \tau ,{v_{{\rm{free}}}},v_s^t))
\end{equation}

In Eq.\ref{8}, \(rand\) represents a random number between [0,1] and \({a}'\) is the random acceleration of the target vehicle. When the velocity of the target vehicle satisfies \(v_{tar}^{t}\ne \text{0}\), the random acceleration is the minimum acceleration, which is because the noise is small when the vehicle is in motion.

• Location update.
\begin{equation}\label{10}
x_{tar}^{t{\rm{ + }}1} = x_{tar}^t + v_{tar}^{t{\rm{ + }}1} \cdot \tau 
\end{equation}

The car-following model involves many variables, for the reader's convenience, as they are used throughout the paper. The setting of related parameters is shown in Tab.\ref{table2}.

\begin{table}[h]
\centering
\caption{Corresponding to different free flow velocities, model parameters and characteristic values.}
\begin{tabular}{*{12}{c}}  
\toprule
\({{v}_{free}}\) & \({{v}_{p}}\) & \({{a}_{\min }}\) & \({{a}_{\max }}\) & \(\lambda \) & \({{k}_{1}}\) & \({{k}_{2}}\) & \({{p}_{0}}\) & \({{p}_{1}}\) & \({{p}_{2}}\) & \({{p}_{a1}}\) & \({{p}_{a2}}\) \\
\midrule
30.24 km/h (21\(\cdot\delta v\)) & 10\(\cdot\delta v\) & \(\delta a\) & 3\(\cdot\delta a\) & 1/4 & 3.55 & 2.2 & 0.4 & 0 & 0.06 & 0.08 & 0.052 \\
40.32 km/h (28\(\cdot\delta v\)) & 13\(\cdot\delta v\) & \(\delta a\) & 4\(\cdot\delta a\) & 1/5 & 3.65 & 2.2 & 0.4 & 0 & 0.06 & 0.08 & 0.052 \\
50.40 km/h (35\(\cdot\delta v\)) & 17\(\cdot\delta v\) & \(\delta a\) & 4\(\cdot\delta a\) & 1/5 & 3.75 & 2.1 & 0.4 & 0 & 0.06 & 0.08 & 0.052 \\
60.48 km/h (42\(\cdot\delta v\)) & 20\(\cdot\delta v\) & \(\delta a\) & 5\(\cdot\delta a\) & 1/6 & 3.85 & 2.1 & 0.4 & 0 & 0.06 & 0.08 & 0.052 \\
\bottomrule
\end{tabular}
\label{table2}
\end{table}

\subsubsection{Lane-changing model} \label{subsection3.1.2}

In a two-lane traffic scenario, the vehicle may change lanes to obtain a better driving condition or to avoid pedestrians. \citet{lv2013modelling} considered the braking distance of vehicles behind adjacent lanes as a safety condition in its lane-changing rules and proposed a method for calculating lane-changing probability. The rules are appropriately modified and applied to this article, Eqs.\ref{11}-\ref{15} give the specific rules.

\textbf{1. Motivation for lane-changing.} 

In the scenario described in this article, the motivation for the target vehicle to change lanes is to obtain a better driving condition. Therefore, when the target vehicle cannot move in a free-flow state or compare with the current lane, the speed of the leading vehicle in the adjacent lane is faster, or the distance between the vehicles is more significant, the motivation of the target vehicles to obtain a better driving condition is satisfied.

\begin{equation}\label{11}
\left\{ \begin{array}{l}
v_{tar}^t < {v_{free}}\\
{g_{a,front}} > {g_{c,front}}{\rm{ \: or \: }}v_{al}^t > v_l^t
\end{array} \right.
\end{equation}

where \({{g}_{c,front}}\)(\({{g}_{a,front}}\)) denote the bumper-to-bumper space gap between the target vehicle and its leader one in the current (adjacent) lane, \(v_{l}^{t}\) (\(v_{al}^{t}\)) denote the speeds of leader one in the current (adjacent) lane.

\textbf{2. Safety conditions for lane-changing.}  

When the target vehicle changes lanes, the distance between the leader one (vehicle or pedestrian) and the follower vehicle in the target lane must be considered. If the distance is small, the requirements for safe lane change will not be met. Therefore, the following relationships need to be satisfied.

\begin{equation}\label{12}
{g_{a,back}} \ge {S_{ab}}{\rm{\: \& \:}}{g_{a,front}} \ge {S_{tar}}
\end{equation}

\begin{equation}\label{13}
\left\{ \begin{array}{l}
S_{ab} = \lceil v_{ab}^t \cdot t_{re} + {v_{ab}^{t^2}}/{2}  \mu   g \rceil\\
S_{tar} = \lceil v_{tar}^t \cdot t_{re} + {v_{tar}^{t^2}}/{2}  \mu   g \rceil
\end{array} \right.
\end{equation}

In Eq.\ref{12}, \({{g}_{a,back}}\) denotes the bumper-to-bumper space gap between the target vehicle and the follower vehicle in the adjacent lane. In Eq.\ref{13}, \({{S}_{ab}}\) and \({{S}_{tar}}\) are the safe braking distances of the target vehicle and the follower vehicle in the adjacent lane, respectively, \(v_{ab}^{t}\) is the speed of the follower one (vehicle or pedestrian) in the adjacent lane at \(t\)-th timestep, \({{t}_{re}}\) is the driver's reaction time, \(\mu \) is the static friction coefficient when the vehicle brakes, and \(g\) is the acceleration of gravity.

\textbf{3. Probability for lane-changing.}

When the above two conditions are met, the target vehicle can change lanes. However, due to the randomness of the driver's strategy and different road conditions, different willingness to change lanes exists.

\begin{equation}\label{14}
{p_c} = \alpha  \cdot {g_{a,front}}/({g_{a,front}} + {g_{c,front}}) + (1 - \alpha ) \cdot v_{al}^t/(v_{al}^t + v_l^t)
\end{equation}

Where, \({{p}_{c}}\) is the probability of lane-changing, and \(\alpha \) is the weight coefficient, \(v_{al}^{t}\) is the speed of the leader one (vehicle or pedestrian) in the adjacent lane at \(t\)-th timestep.

\textbf{4. Execution for lane-changing.}  

When the above three conditions are met, the target vehicle will change lanes. Due to the discrete nature of cellular automata, we assume that the lane-changing behavior is completed within one timestep.

\begin{equation}\label{15}
\left\{ \begin{array}{l}
flag = 1{\rm{        \quad     }}rand \le {p_c}\\
flag = 0{\rm{        \quad   }}{p_c} < rand \le 1
\end{array} \right.
\end{equation}

\begin{figure}[ht!]
\centering
\includegraphics[scale=0.3]{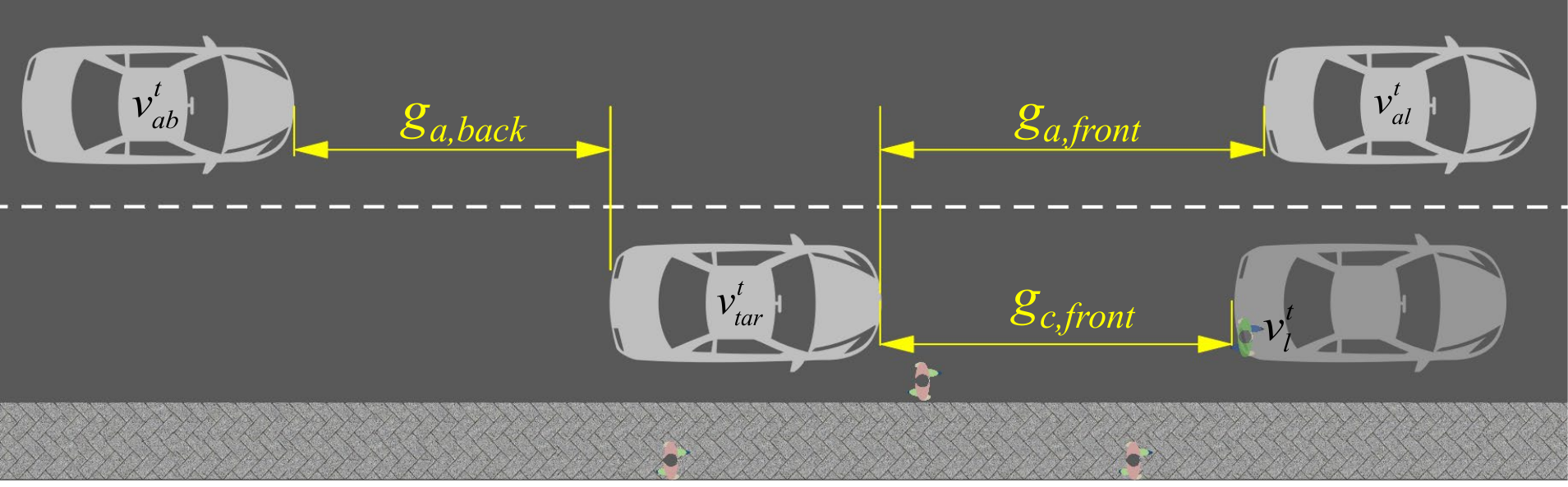}
\caption{Schematic diagram of relevant variables when the vehicle changes lanes.}
\label{fig6}
\end{figure}

Where, \(flag=1\) means changing lanes, \(flag=0\) means not changing lanes. The explanation of symbols is shown in Fig.\ref{fig6}. The lane-changing model involves additional variables to those of the car-following model, the setting of related parameters is shown in Tab.\ref{table3}.

\begin{table}[h]
\centering
\caption{Model parameters and characteristic values.}
\begin{tabular}{ccccc}
\hline
Parameters & \({{t}_{re}}\) & \(\mu\) & \(g\) & \(\alpha\) \\
\hline
Value & 0.4 (\(s\)) & 0.4 & 10 (m/s\(^2\)) & 0.4 \\
\hline
\end{tabular}
\label{table3}
\end{table}

\subsubsection{Analysis of the IKKW model} \label{subsection3.1.3}

\begin{figure}[ht!]
\centering
\includegraphics[scale=0.5]{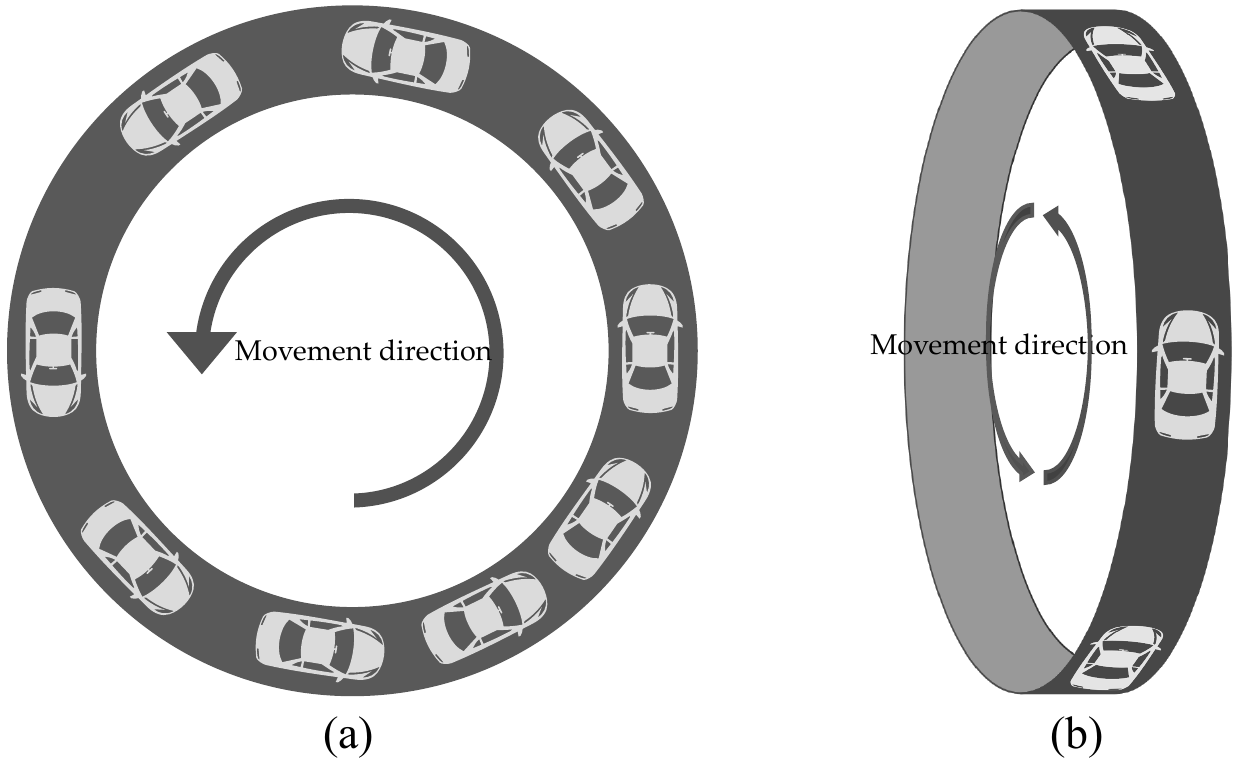}
\caption{ Two-dimensional (a) and three-dimensional (b) schematic diagrams of periodic boundary conditions.}
\label{fig7}
\end{figure}

Whether the vehicle movement model can simulate the mixed traffic scene well, it depends on its accurate description of the microscopic motion characteristics of the vehicle. The mixed traffic simulation model needs to model accurate pedestrian-vehicle conflicts, and the design of vehicle deceleration rules is the model's focus.
In the analysis part of the model, a section of 800m single-lane road was considered to carry out the comparative analysis of the IKKW model and the KKW-1 model in terms of the car-following behavior. The road is set with a periodic boundary condition. The vehicle will enter the road from the beginning of the road at the same time as another one leaves the road from the end of the road, so the periodic boundary road can be regarded as a circular road in which it’s beginning and end are connected in the two-dimensional plane (as shown in Fig.\ref{fig7}(a)), and in the three-dimensional space, the periodic boundary road can be regarded as a ring in space (as shown in Fig.\ref{fig7}(b)).

The cell size of both models is set as 0.4 m \(\cdot\) 0.4 m, and the vehicle length is set as 6 m. The free flow speed is set as 60.48 km/h. Other parameter settings of the IKKW model are shown in Tab. \ref{table2}, and the parameter settings of the KKW-1 model are consistent with the literature \citep{kerner2002cellular}.

The simulations were carried out in the synchronized flow (\({{\rho }_{veh}}\)=65 veh/km) and blocked flow (\({{\rho }_{veh}}\)=100 veh/km) environments, and the two models were simulated independently, so that a total of 4 sets of simulations were performed. In each set of simulations, the first 1000 timesteps are discarded to let the transient time out and the data was collected for the next 200 timesteps. The comparison of the micro-car-following behavior in the two models is analyzed by extracting the trajectories and related motion parameters of five vehicles (sequentially numbered car 1-5) over the 200 timesteps.

Fig.\ref{fig8} and Fig.\ref{fig9} show the time-space diagram for both models and the corresponding vehicles' displacement, speed and acceleration variations in the synchronized and blocked flows. By comparing with the KKW-1 model, it can be seen that in the synchronized flow, the IKKW model has almost the same model performance as the KKW-1 model, and the vehicle-to-vehicle exhibits a strong speed adaptation phenomenon with vehicles’ speed and acceleration fluctuating in a small range. However, in the blocked flow, the performance of the two models showed a large difference. Due to the constant acceleration setting in the KKW-1 model, there is a long acceleration period when the vehicle starts from a standstill, and the vehicle suddenly comes to a standstill without decelerating in time due to the constant synchronization distance setting as the vehicle approaches the blockage area. In Fig.\ref{fig9}(d), it can be seen that the KKW-1 model exhibits extremely drastic acceleration changes. In contrast, in the IKKW model, the vehicle follows the relaxation behavior naturally and smoothly without extreme and unnatural motion behavior during acceleration and deceleration due to the dynamic acceleration and synchronization distance settings.

\begin{figure}[ht!]
\centering
\includegraphics[scale=0.85]{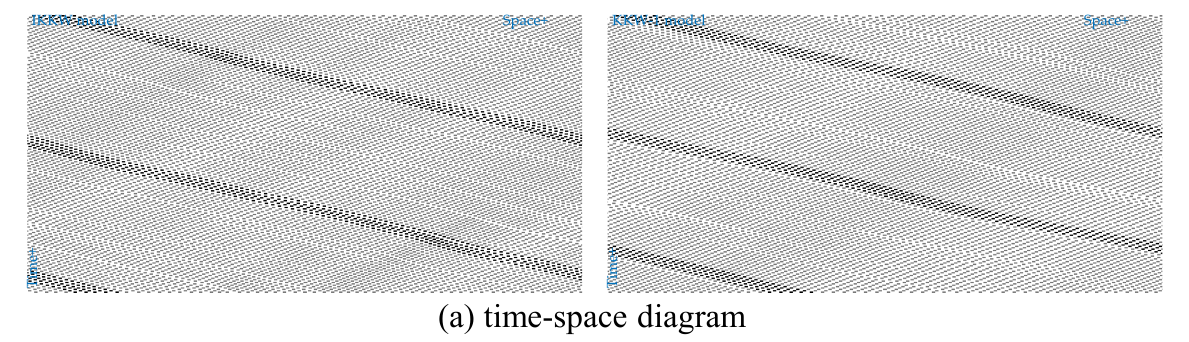}
\caption{Fig. 8. Time-space diagram and corresponding displacement, velocity and acceleration change with time in synchronized flow, (\({{\rho }_{veh}}\)=65 {veh/km}).}
\label{fig8}
\end{figure}

\begin{figure}[ht!]\ContinuedFloat
\centering
\includegraphics[scale=0.65]{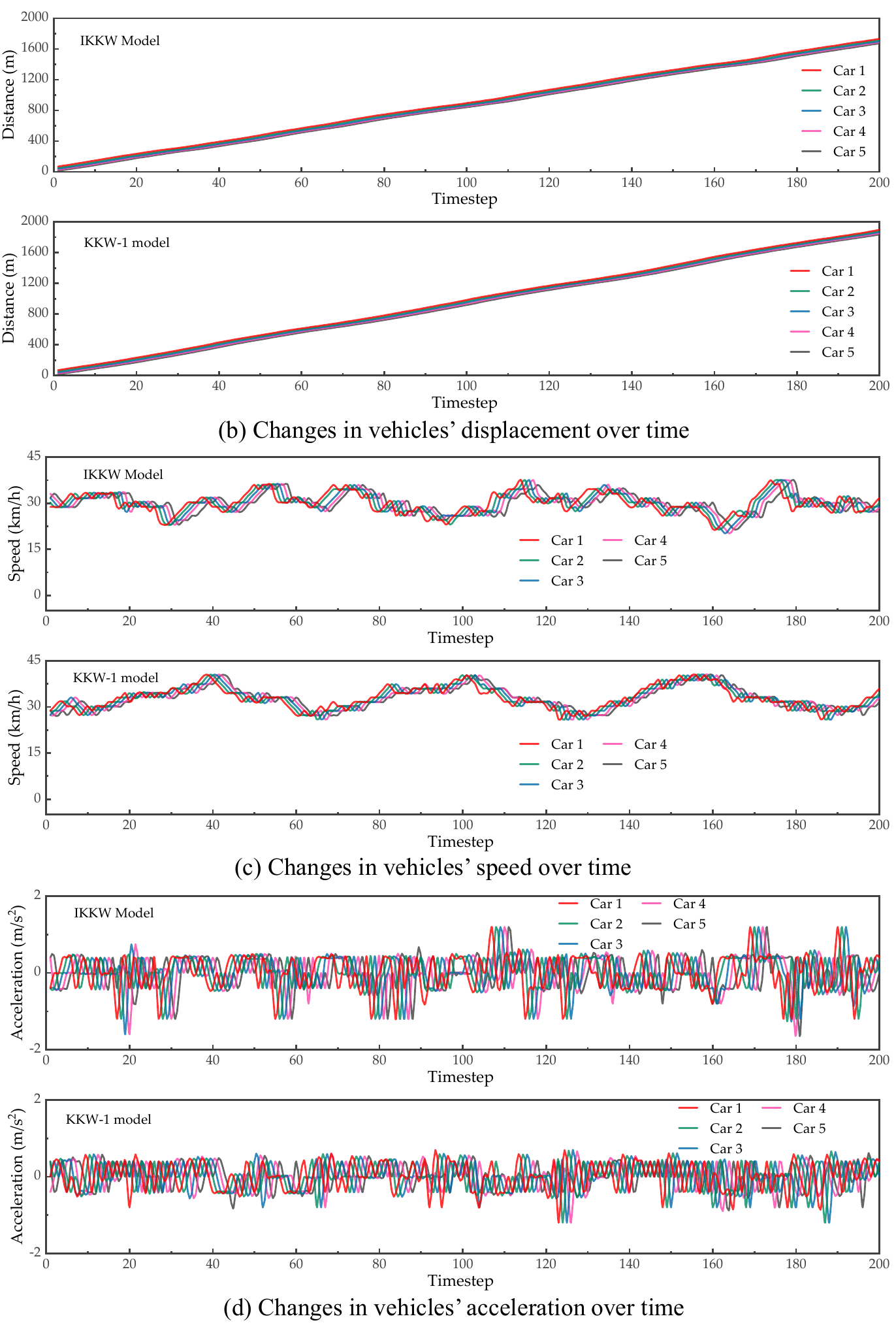}
\caption{Continued.}
\label{fig8c}
\end{figure}

\begin{figure}[ht!]
\centering
\includegraphics[scale=0.85]{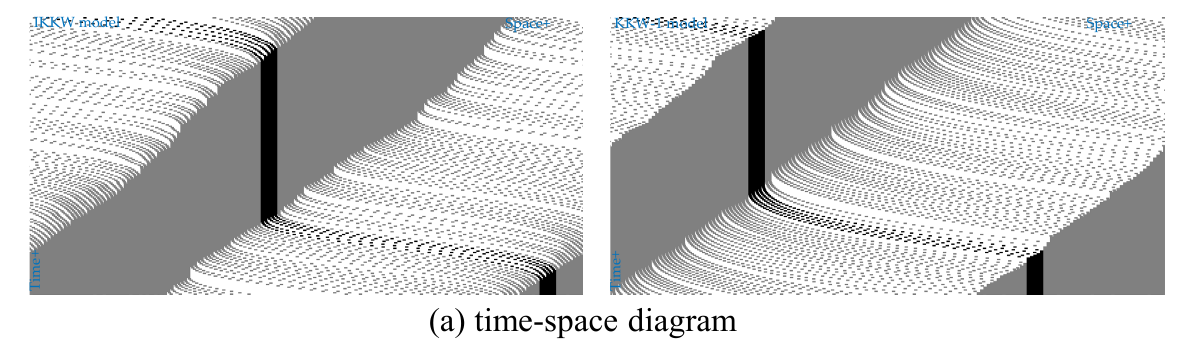}
\caption{Fig. 9. Time-space diagram and corresponding displacement, velocity and acceleration change with time in blocked flow, (\({{\rho }_{veh}}\)=100 veh/km).}
\label{fig9}
\end{figure}

\begin{figure}[ht!]\ContinuedFloat
\centering
\includegraphics[scale=0.65]{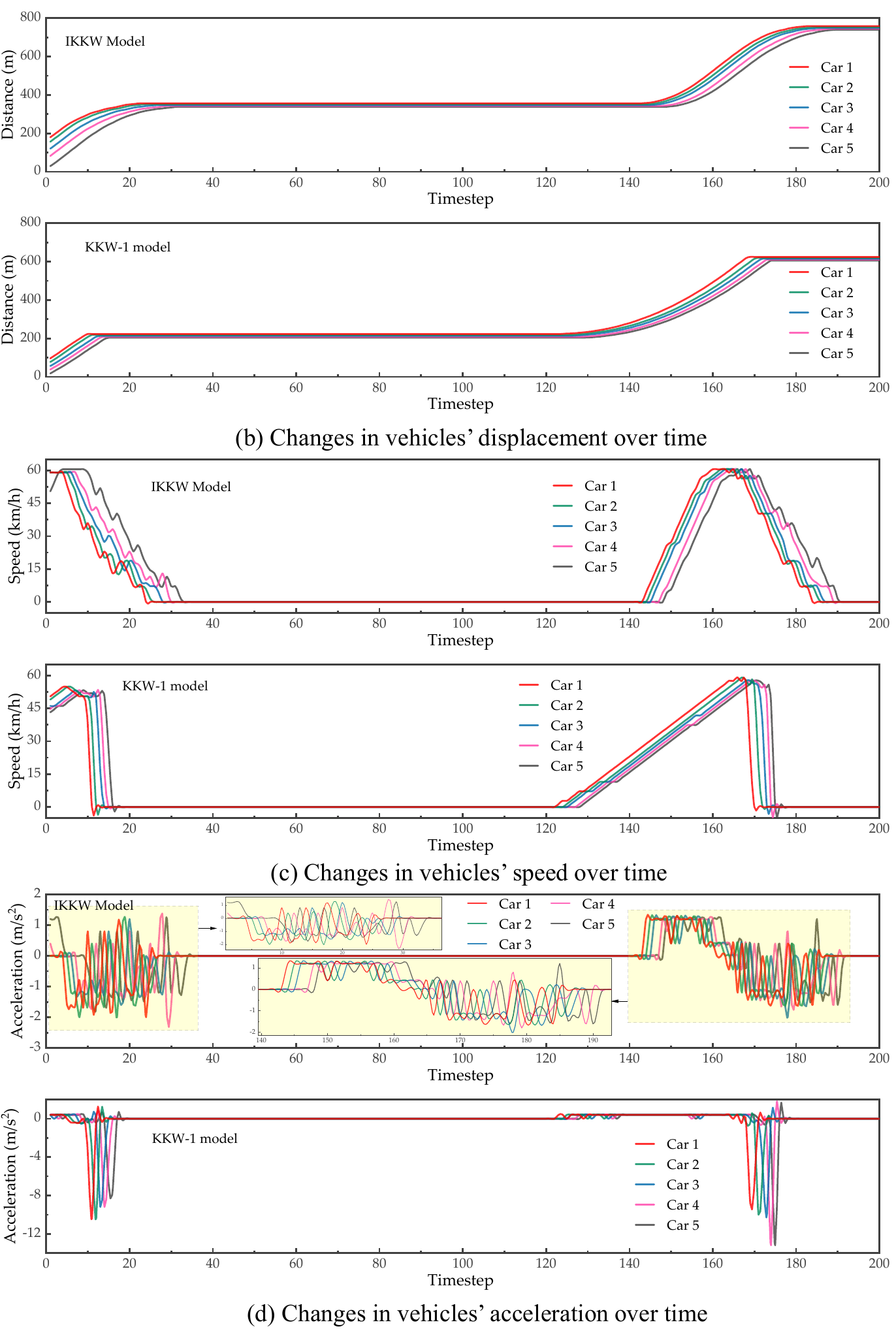}
\caption{Continued.}
\label{fig9c}
\end{figure}

This model can accurately reflect the characteristics of vehicle deceleration, speed adaptation, and car-following motion. The speed and acceleration fluctuation of the blocked flow is stronger than that of the synchronized flow state. The IKKW model performs better than the KKW-1 model in terms of the accuracy of the car-following behavior.

\subsection{Pedestrian movement model} \label{subsection3.2}

In the field of pedestrian flow, a large number of models have studied the movement mechanism, such as the pedestrian movement strategy based on the nearest neighbor method \citep{ma2010k,zhao2021radar}, turning behavior of pedestrians in a bidirectional pedestrian flow \citep{yamamoto2019body}, pedestrian conflict resolution mechanism based on game theory \citep{asano2010microscopic}, the pedestrian model of different view angle discretization \citep{asano2010microscopic,liu2014agent} and field of view \citep{lv2013two}, what's more, artificial neural network methods have also been applied to pedestrian motion modelling \citep{zhao2021radar,ma2016artificial}. These works have further promoted the development of pedestrian flow research. In terms of pedestrian flow modeling using cellular automata, early scholars established a random walker model without step back to simulate pedestrian motion behavior in the passage \citep{yu2007cellular} and proposed an asynchronous update model to simulate the multi-velocity nature of pedestrians \citep{weng2006cellular,li2020relationship}, by observing pedestrians' walking preferences, the multi-field based model was proposed \citep{nowak2012quantitative,yue2010study}. \citet{ma2010k} pointed out that \(k\)-Nearest-Neighbor (\(k\)-NN) interaction plays the fundamental role in the emergence of collective pedestrian phenomena, \citet{xue2016behaviour} considered the psychological decision-making of pedestrians in the bidirectional pedestrian flow, and the concept of a dominant row was introduced to model the lane-formation phenomenon.

\subsubsection{Model rules} \label{subsection3.2.1}
In the research scenario of this article, the pedestrian movement on the sidewalk is more similar to unidirectional walking, the sidewalk and the motorway are often separated by a road shoulder. Unlike hard isolation barriers such as walls, the setting of road shoulders is regarded as a soft isolation measure and cannot prevent the intrusion of pedestrians well. Similar to Blue's model \citep{blue2001cellular}, we divide the pedestrian's movement into two parts: lateral movement and longitudinal movement, similar to the lane changing and car-following behaviors in the vehicle model. Pedestrians need to move laterally to complete overtaking and avoiding behaviors. In addition, the risk floor field is introduced. When the pedestrian is on the sidewalk, the value of the risk floor field is 0, which means there is no danger of colliding with the vehicle. In lane 2, the dangerous field increased from 0.3 to 0.9, indicating that the longer the depth of the pedestrian intrusion, the greater is the corresponding risk. In lane 1, the value of the risk floor field is always 1, indicating an absolutely dangerous state, which pedestrians cannot enter. In addition, the Time-To-Collision (TTC) has also been considered.

The rules of pedestrian movement are set as follows:

\textbf{1. Pedestrian lateral movement.}

• Calculate the time-to-collision (TTC) of the target pedestrian (\({{t}_{TTC}}\)).

\begin{equation}\label{16}
{t_{TTC}} = \left\{ \begin{array}{l}
{\rm{ }}{g_{back}}/(v_b^t - v_{tar}^t){\rm{ \quad if\: }}v_b^t > v_{tar}^t\\
{\rm{ }}{g_{back}}/v_b^t{\rm{  \;\;\:   \quad \quad  \qquad if\: }}v_b^t \le v_{tar}^t{\rm{\: and\: }}{v_{b,n}} \ne 0\\
\;len{\rm{ }}/{v_{free}}{\rm{   \qquad  \qquad if\: no\: car\: behind \:or \:}}v_b^t = 0
\end{array} \right.
\end{equation}

Where, \({{g}_{back}}\) denotes the distance between the target pedestrian and the rear vehicle, \(v_{b}^{t}\) denotes the vehicle's speed behind the target pedestrian at \(t\)-th timestep, and \(len\) denotes the length of the simulated road (corresponding to 1250\(\cdot\delta x\) in this paper). In the third term of the equation, pedestrians are still assigned a small TTC value even if they are located on the sidewalk (corresponding to the case of no car behind), this is considering that pedestrians walking on the sidewalk will be influenced by vehicles and choose to walk away from them even if there is no car behind them, which makes the simulated cumulative density distribution of pedestrians appear to walk with a significant rightward tendency, as shown in Fig.\ref{fig11}.

• Calculate the probability of direction selection (turning left or keeping stationary or turning right)

Suppose the target pedestrian movement area is abstracted as a motion space consisting of channels. In that case, target pedestrian always expects to choose the channel with less risk and a more open view ahead (greater spacing from pedestrians ahead), so in the model, the pedestrian's choice for the lateral direction depends on the risk field (corresponding to the magnitude of risk) and the space not occupied ahead. The probability is calculated as:

\begin{equation}\label{17}
\left\{ \begin{array}{l}
{p_l} = \beta  \cdot (1 - {r_l}) + \gamma  \cdot \min ({g_{l,front}},\varphi  \cdot {v_{\max }})\\
{p_c} = \beta  \cdot (1 - {r_c}) + \gamma  \cdot \min ({g_{front}},\varphi  \cdot {v_{\max }})\\
{p_r} = \beta  \cdot (1 - {r_r}) + \gamma  \cdot \min ({g_{r,front}},\varphi  \cdot {v_{\max }}) + 1/{t_{TTC}}
\end{array} \right.
\end{equation}

In Eq.\ref{17}, \({{r}_{l}}\), \({{r}_{c}}\) and \({{r}_{r}}\) denote the value of the risk field corresponding to the left, current position, and right of the target pedestrian, respectively, is a dimensionless parameter with a range of [0,1] (\({{r}_{c}}\ne 0\) indicates that the pedestrian invades the motorway). \({{v}_{\max }}\) denotes the maximum speed of the target pedestrian in the longitudinal direction as a constant, \({{p}_{l}}\), \({{p}_{c}}\)and \({{p}_{r}}\) denote the probability of the pedestrian turning left, keeping stationary and turning right, respectively, and \({{g}_{l,front}}\), \({{g}_{front}}\) and \({{g}_{r,front}}\) denote the space gap in front of the left, in front ahead and in front of the right of the target pedestrian, respectively. In addition, \(\beta \), \(\gamma \) and \(\varphi \) denote the corresponding weights as dimensionless parameters, where \(\varphi \cdot {{v}_{\max }}\) denotes the maximum space gap ahead that affects the lateral direction choice of the target pedestrian.

• Calculate the target pedestrian movement probability (\({{p}_{move}}\)).

\begin{equation}\label{18}
{p_{move}} = \left\{ \begin{array}{l}
{p_m}{\rm{                                                         \qquad\qquad\qquad\qquad\quad\qquad\qquad\qquad\qquad\;\; if\: }}{r_c} \ne 0\\
({v_{\max }} - v_{tar}^t{\rm{)(}}{p_l} + {p_r})/(\sum\nolimits_{i = l,c,r} {{p_i}}  \cdot {v_{\max }}{\rm{) \quad if \:}}{r_c} = 0
\end{array} \right.
\end{equation}

The first term in the Eq.\ref{18} indicates the lateral movement with a fixed probability (\({{p}_{m}}\)) when the target pedestrian is located in the motorway. The second term indicates the movement probability when the pedestrian is in the crosswalk. The smaller the target pedestrian's speed, the more he/she expects to obtain a favorable position to increase his/her speed through lateral motion. The movement probabilities are normalized so that \({{p}_{move}}\) always lies in the interval [0,1].

• Lateral movement

Lateral movement according to the above direction selection probability and movement probability (\(ran{{d}_{1}}\ne ran{{d}_{2}}\)).

\begin{equation}\label{19}
\left\{ \begin{array}{l}
fla{g_l}{\rm{ = 1 \qquad\qquad\qquad \, if \: }}ran{d_1} < {p_l}{\rm{/}}\sum\nolimits_{i = l,c,r} {{p_i}} {\rm{\: and \:}}ran{d_2} < {p_{move}}\\
fla{g_r}{\rm{ = 1        \qquad\qquad\qquad      if \: }}{p_l}{\rm{/}}\sum\nolimits_{i = l,c,r} {{p_i}}  \le ran{d_1} < ({p_l} + {p_r}{\rm{)/}}\sum\nolimits_{i = l,c,r} {{p_i}} {\rm{ \: and \: }}ran{d_2} < {p_{move}}\\
fla{g_l} = 0\:\&\: fla{g_r} = 0{\rm{  \; \:   otherwise}}
\end{array} \right.
\end{equation}

When pedestrians move laterally, one timestep can only move at most one cell (\(\delta x\)). Where, \(fla{{g}_{l}}\text{=1}\) means pedestrians move to the left, \(fla{{g}_{r}}=1\) means pedestrians move to the right.

\textbf{2. Pedestrian longitudinal movement.}

• Longitudinal speed update.

\begin{equation}\label{20}
v_{tar}^{t+1/2}=\min (v_{tar}^{t}+1,{{g}_{front}},{{v}_{\max }})
\end{equation}

\begin{equation}\label{21}
v_{tar}^{t\text{+}1}=v_{tar}^{t\text{+}1/2}-1\text{   if }rand<{{p}_{slow}}
\end{equation}

Eq.\ref{20} describes the acceleration process of the target pedestrian, and Eq.\ref{21} represents the random deceleration behavior of the pedestrian. \({{p}_{slow}}\) is the random slowing probability of the pedestrian with a dimensionless constant.

• Location update.

\begin{equation}\label{22}
x_{tar}^{t+1}=x_{tar}^{t}+v_{tar}^{t\text{+}1}\cdot\tau 
\end{equation}

\begin{figure}[ht!]
\centering
\includegraphics[scale=0.5]{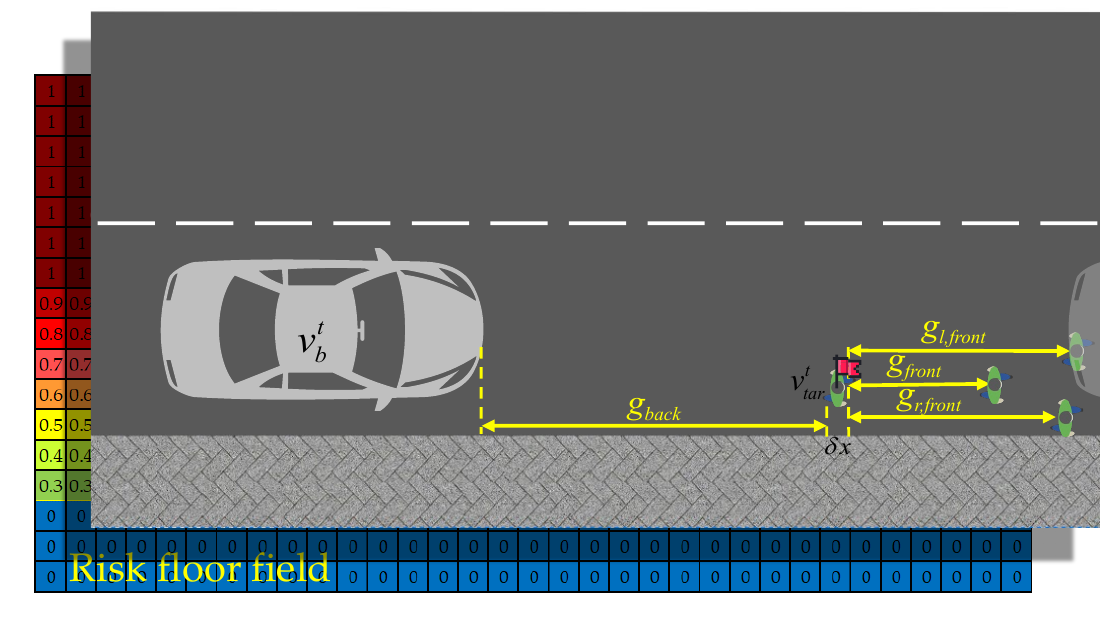}
\caption{Schematic diagram of pedestrian movement.}
\label{fig10}
\end{figure}

The explanation of variables is shown in Fig.\ref{fig10} and  the setting of related parameters is shown in Tab.\ref{table4}.

\begin{table}[h]
\centering
\caption{Model parameters and characteristic values.}
\begin{tabular}{ccccccc}
\toprule
Parameters & \(p_{\text{slow}}\) & \(\beta\) & \(\gamma\) & \(\varphi\) & \(v_{\max}\) & \(p_m\) \\
\midrule
Value & 0.1 & 0.5 & 0.05 & 3 & \(3\cdot\delta v\) & 0.8 \\
\bottomrule
\end{tabular}
\label{table4}
\end{table}

\subsubsection{Validation} \label{subsection3.2.2}
\textbf{1. Pedestrian movement characteristics.}

\begin{figure}[ht!]
\centering
\includegraphics[scale=0.6]{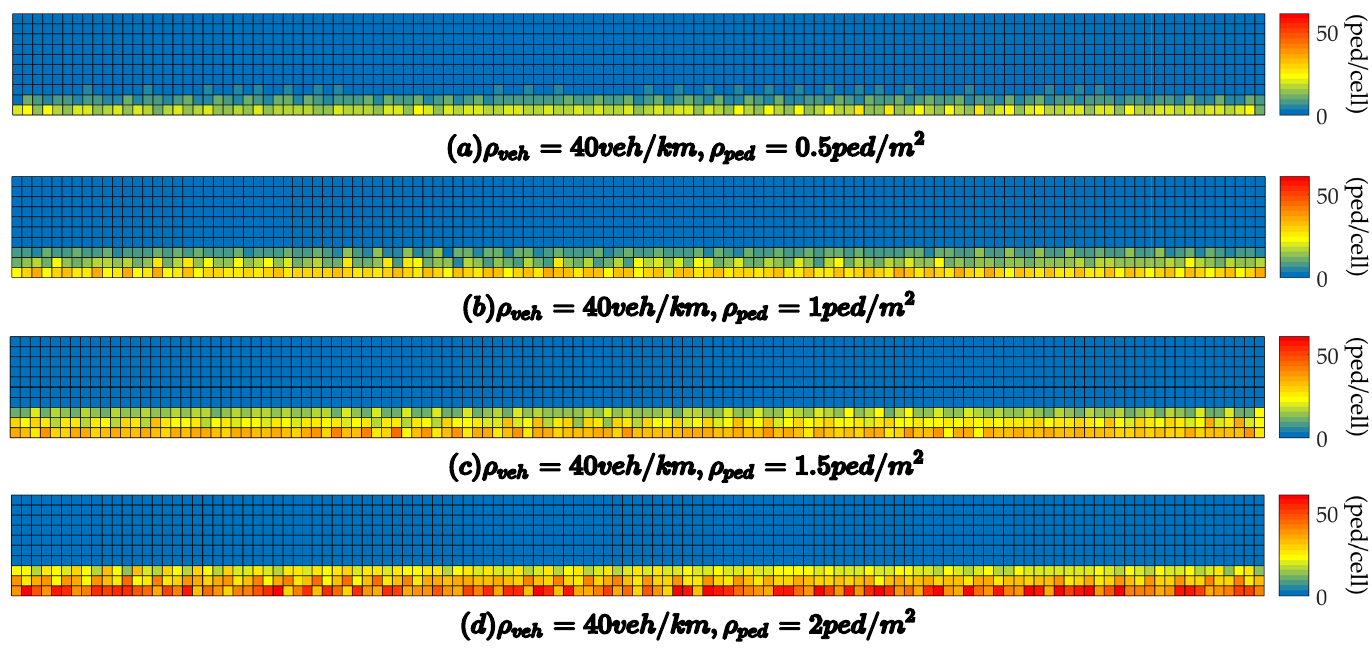}
\caption{Cumulative pedestrians’ density distribution heatmap.}
\label{fig11}
\end{figure}

\begin{figure}[ht!]
\centering
\includegraphics[scale=0.6]{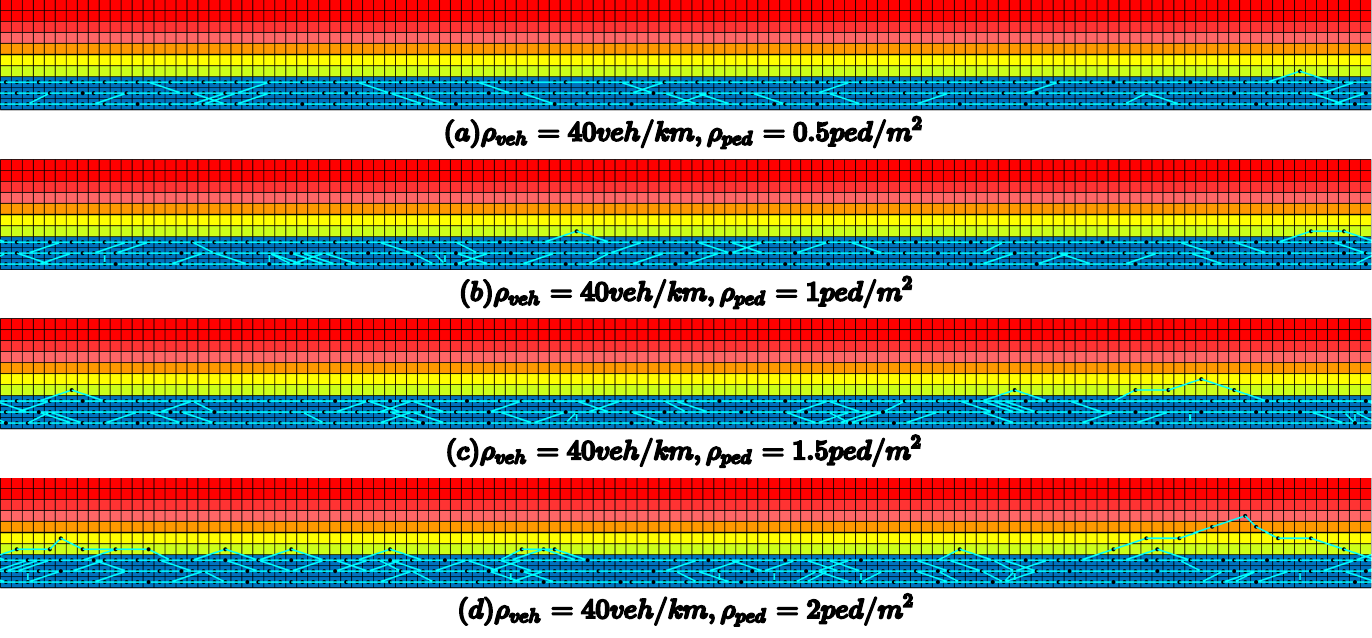}
\caption{Cumulative pedestrian space trajectory map (the colour of the cell indicates the size of the risk value, corresponding to Fig.\ref{fig10}, a higher colour temperature means higher risk).}
\label{fig12}
\end{figure}

The simulation is carried out for a cellular space with a periodical boundary, representing a circular road to observe the microscopic characteristics of pedestrian movement in the model. The road scene is set to a motorway adjacent to the sidewalk, and the road length and width are set as 50 m and 4 m, respectively. In the simulation experiments the vehicles density is set to 40 veh/km/lane, and the initial density of pedestrians on the sidewalk is set  to 0.5 ped/m\(^2\), 1 ped/m\(^2\), 1.5 ped/m\(^2\) and 2 ped/m\(^2\), respectively. The relevant parameter settings are shown in Tab.\ref{table2}, Tab.\ref{table3} and Tab.\ref{table4}.
In each set of simulations, the first 1000 timesteps are discarded to not account for the transient time and the data are collected for the next 100 timesteps. The cumulative pedestrians’ density distribution heatmap (Fig.\ref{fig11}) and cumulative spatial trajectory map (Fig.\ref{fig12}) of 100 timesteps were obtained.

%%%%%%%%%

It can be seen from Fig.\ref{fig11} and Fig.\ref{fig12} that when the pedestrians’ density is low, pedestrians are more inclined to walk on the sidewalk, as the density increases, the frequency of pedestrian intrusion into the motorway begin increase due to the restricted space for movement. From Fig.\ref{fig11}, it can be seen that since the TTC is incorporated into the pedestrian movement rules, even on the sidewalk, pedestrians are more inclined to walk on the right.

\textbf{2. Fundamental diagram.}

\begin{figure}[ht!]
\centering
\includegraphics[scale=0.55]{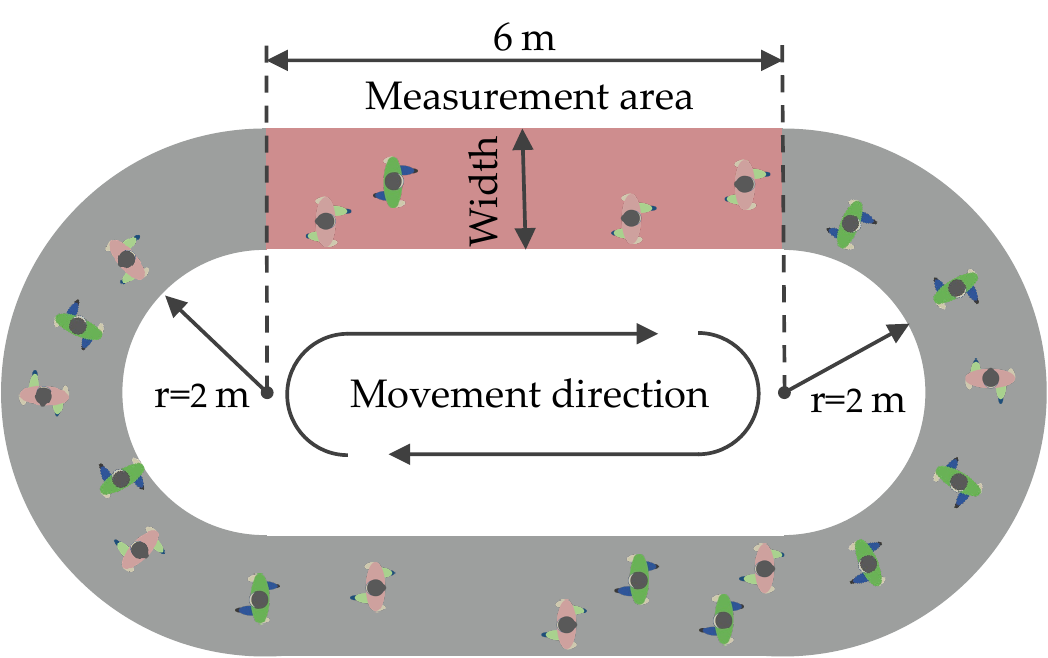}
\caption{Illustration of the experiment scenario.}
\label{fig13}
\end{figure}

\begin{figure}[ht!]
\centering
\includegraphics[scale=0.6]{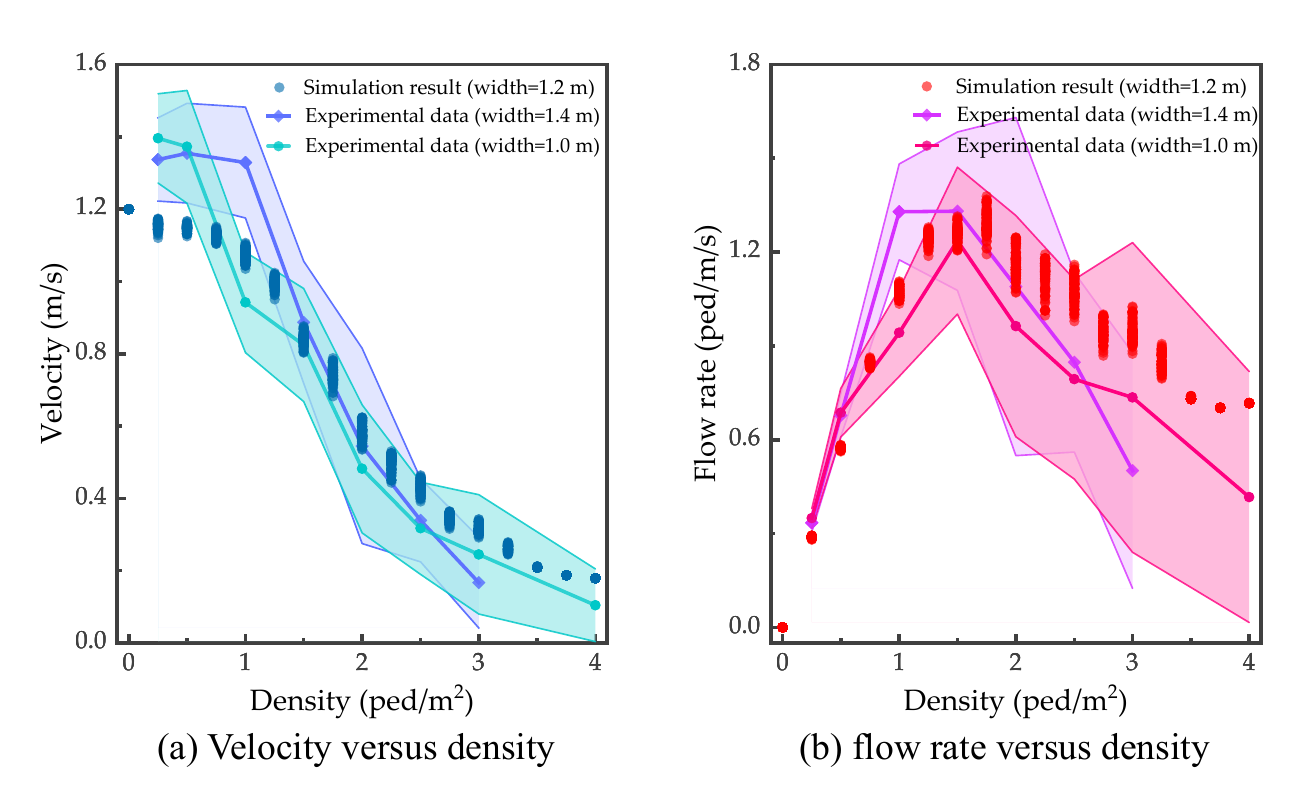}
\caption{Comparison of fundamental diagrams of experimental and model data. The vertical intervals in the figure represent the changing trend of the data mean: the shaded areas represents one-sigma error(mean±standard deviation). Source: The data were taken from \url{https://ped.fz-juelich.de/da/doku.php}.}
\label{fig14}
\end{figure}

To analyze the fundamental diagram of the pedestrian movement model, we do not consider pedestrian intrusion into the lane to establish a simulation scenario in which pedestrians can only walk on the sidewalk. The simulation scenario is set up as a periodic boundary condition sidewalk with a width of 1.2 m and a length of 500 m, so pedestrian intrusion into the motorway is not considered. With [0, 4] as the pedestrians’ density range and  \({\rho _{ped}} = 0.25{\rm{ ped/}}{{\rm{m}}^2}\) as the interval, a total of 15 simulations were carried out (Simulations of \({\rho _{ped}} = 0 {\rm{ ped/}}{{\rm{m}}^2}\)  are excluded). In each set of simulations, the first 1000 timesteps are discarded to not account for the transient time and the data are collected for the next 500 timesteps. For these values of the parameters, data of the average speed and average flow rates of pedestrians are extracted at intervals of every 5 timesteps.
Correspondingly, this paper uses the experiment data of unidirectional pedestrian flow obtained from the database of the Institute for Advanced Simulation (IAS) to verify the model. The experimental scenario is shown in Fig.\ref{fig13}, and the experimental data come from \citep{keip2009dokumentation}. The pedestrian movement data under different channel widths were obtained. This paper draws the fundamental diagram of unidirectional pedestrian flow by extracting the pedestrian velocity and flow rate data under the channel width conditions of 1.0 m and 1.4 m, as shown in Fig.\ref{fig14}. The measurement method of all simulation and experimental data adopted method C of reference \citep{zhang2011transitions}.
By comparing the fundamental diagrams of the simulation data and the experimental data (as shown in Fig.\ref{fig14}), it can be seen that the fundamental diagrams of this model and the experiment have high consistency in the changing trends. The velocity distribution of this model is lower than of the experimental data when the density is lower due to the constant maximum velocity of 1.2 m/s set in this paper.

\subsection{Model assumption} \label{subsection3.3}

In actual traffic scenarios, the motion of pedestrians and vehicles is very diverse, this significantly increases calculations of the model. To simplify this model, the following assumptions are made:

•  Vehicles always drive in the middle of the lane, and moderate intrusion behavior will not affect the vehicle. (In this model, when the pedestrian intrusion depth is 1 cell , the vehicle is considered to be able to drive without adjusting its course for pedestrians. When the invasion depth exceeds 1 cell, the vehicle needs to evade.)

•  Invading pedestrians can only move in lane 2 and they cannot enter lane 1.

•  When a pedestrian is in front, the vehicle only considers two evasive behaviors: slowing down or changing lanes (the behavior of vehicles circumventing to avoid pedestrians is not considered).

\begin{figure}[ht!]
\centering
\includegraphics[scale=0.6]{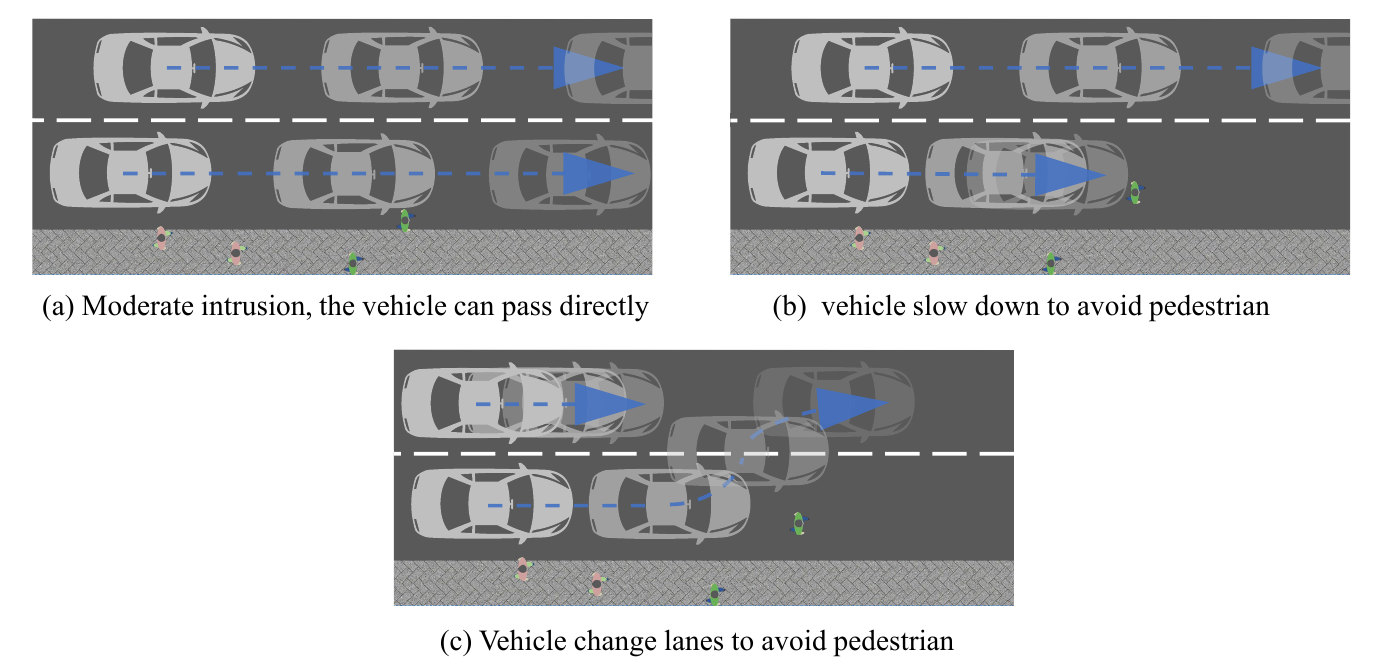}
\caption{Vehicle avoidance behavior.}
\label{fig15}
\end{figure}

In a mixed traffic scenario, considering that vehicles give way to pedestrians, pedestrians have the priority of movement and decision-making. Therefore, when there is a conflict between a pedestrian and a vehicle, the update of the pedestrian takes precedence over the vehicle. In this way, the pedestrian and vehicle movement models can be well coupled according to the above assumptions, and the related schematic diagram is shown in Fig.\ref{fig15}.

The model was built based on the above rules and settings, which will affect the model's performance to a certain extent. All results of the simulation analysis in the following paper are based on the premise of the assumption above.

\section{Mixed Traffic Simulation\label{section4}}

\begin{figure}[ht!]
\centering
\includegraphics[scale=0.6]{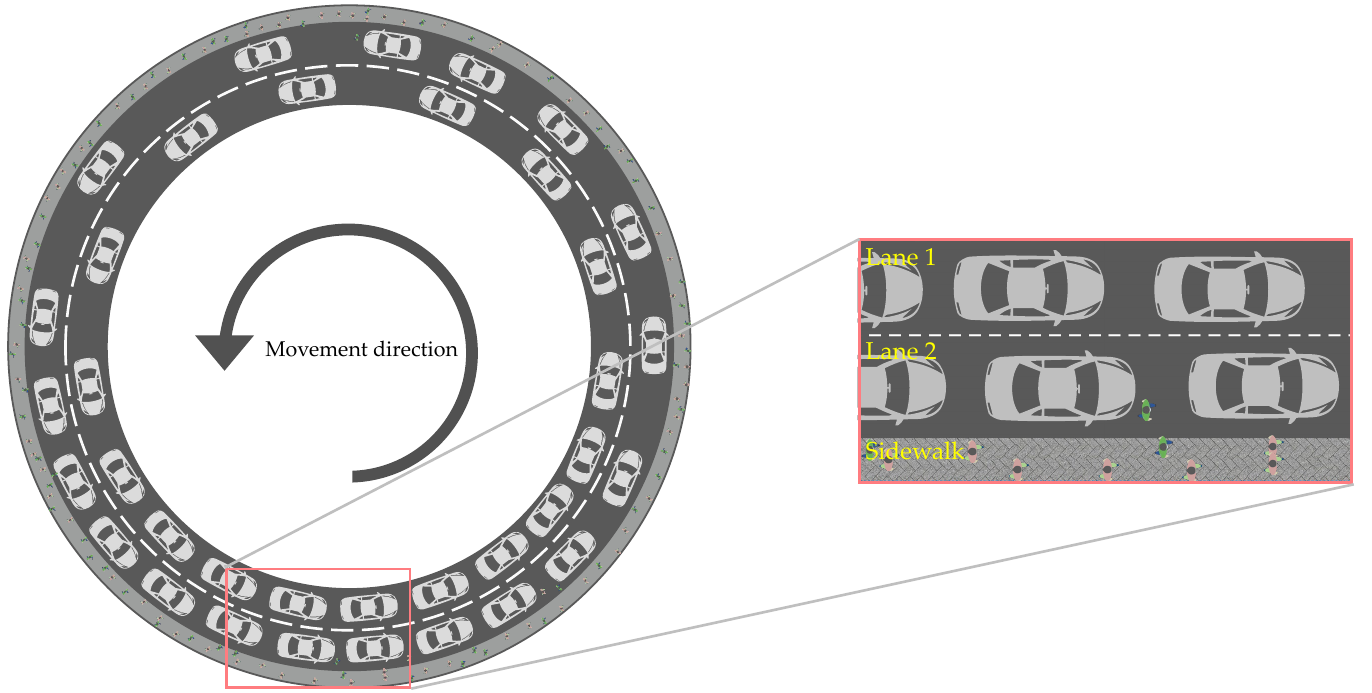}
\caption{Schematic diagram of the simulation scenario.}
\label{fig16}
\end{figure}

The simulated scenario consists of a one-way two-lane motorway with a length of 500 m and a sidewalk adjacent to the motorway, assumes periodic boundary conditions, as shown in Fig.\ref{fig16}. The relevant parameters of the model are shown in Tab. \ref{table2}, Tab. \ref{table3}, and Tab. \ref{table4} (The free flow speed is set as 60.48 km/h). The vehicles’ density was divided into 4 groups at increments of 20 veh/km/h (from 20 veh/km/lane to 80 veh/km/lane). Within each group, the pedestrians’ density is further divided into 9 groups at increments of 0.25 ped/m\(^2\) (from 0 ped/m\(^2\) to 2 ped/m\(^2\)). A total of 36 scenarios were considered and independent simulations were run and obtained traffic time-space maps, speed, flow rate and travel time for each vehicles’ density and pedestrians’ density scenario. In each simulation, the first 1000 timesteps are discarded to let the transient time pass and data were collected for the next 500 timesteps.

\subsection{Undersaturated traffic (\({{\rho }_{veh}}=20 veh/km/lane\))} \label{subsection4.1}

\begin{figure}[ht!]
\centering
\includegraphics[scale=0.75]{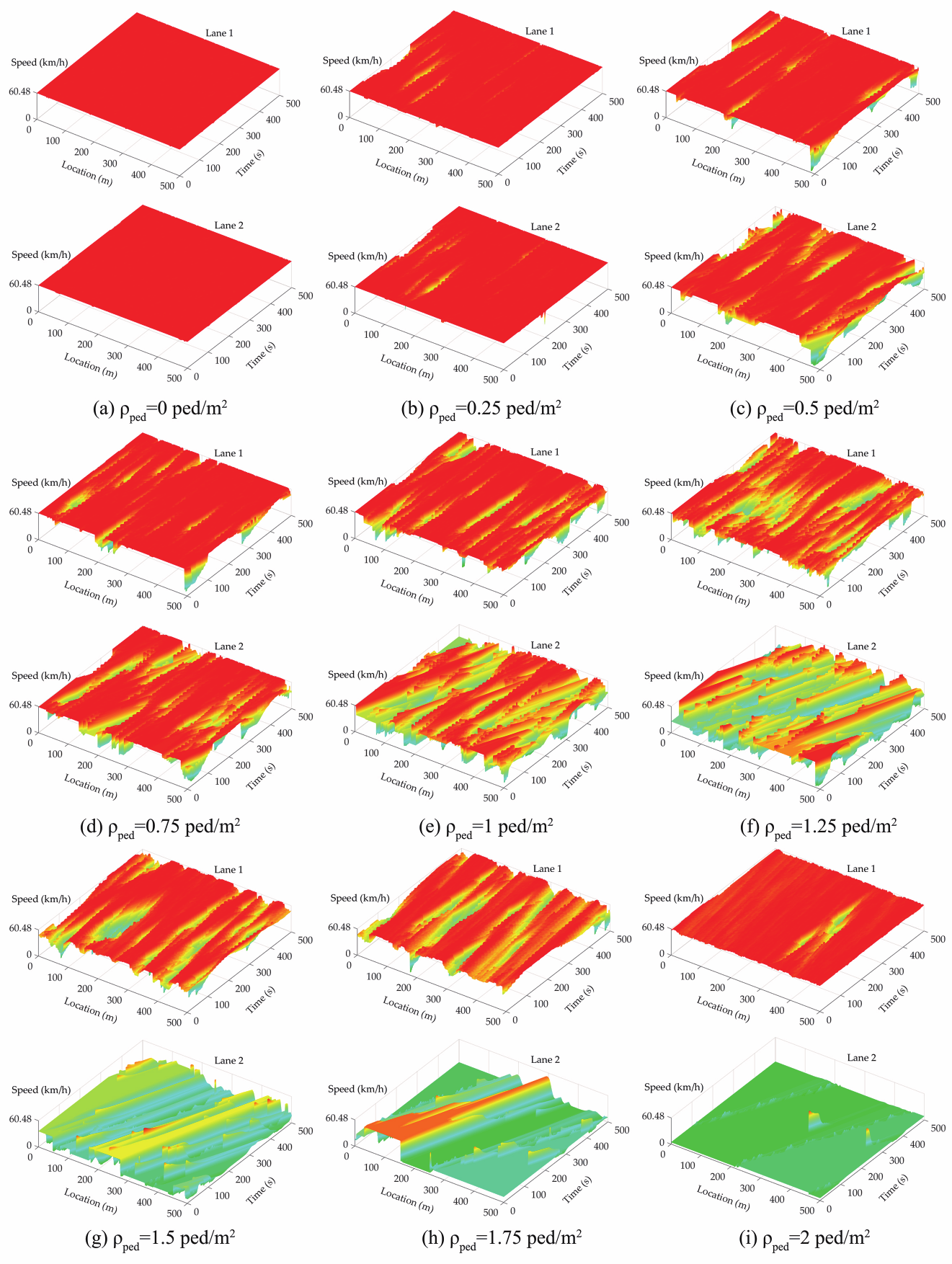}
\caption{Time-space evolution diagram of vehicles’ speed for 500 timesteps, \({{\rho }_{veh}}\)=20 veh/km/lane (We also provide a comprehensive two-dimensional time-space diagram version, please refer to the data availability section).}
\label{fig17}
\end{figure}

In the simulation, the intrusion situation under different pedestrian densities can be observed by setting different initial pedestrian densities on the sidewalk. Fig.\ref{fig17} is the time-space diagram of vehicles’ speed under different pedestrians’ density conditions when the vehicles’ density is 20 veh/km/lane. When the pedestrians’ density is 0 (Fig.\ref{fig17} (a)), the traffic in the simulation scenario is undersaturated, and the vehicles are driving in a free-flow state. With the increase in pedestrians’ density, pedestrian intrusion phenomenon began to appear, and this trend increases with the increase in pedestrians’ density. When the pedestrians’ density rises to 0.25-1 ped/m\(^2\) (Fig.\ref{fig17} (b)-(e)), pedestrians and vehicles faced intensive conflicts. In the time-space diagram, many deceleration behaviors can be observed in lane 2, and the traffic in lane 1 is further affected due to the lane-changing vehicles. When the pedestrians’ density is higher than 1 ped/m\(^2\) (Fig.\ref{fig17}(f)-(i)), most of the vehicles in lane 2 choose to change to lane 1, and there are only a few vehicles travelling in lane 2, further aggravating the intrusion of pedestrians.

Traffic speed and flow rate are the fundamental statistics of the state and characteristics within traffic. The speed and flow rate statistics under different pedestrian densities are analyzed further to study the impact with pedestrian intrusion behaviour on traffic. The average speed \(\bar{v}\), average density \(\bar{\rho }\), average flow rate \(\bar{q}\) of \(n\) vehicles on a single-lane road with a road length of \(L\) in \(T\) timesteps are expressed as:

\begin{equation}\label{22}
\left\{ \begin{array}{l}
\bar v = \sum\limits_{t = 1}^T {\sum\limits_{i = 1}^n {{v_{i,t}}} } /nT\\
\bar \rho  = n/L\\
\bar q = \sum\limits_{t = 1}^T {\sum\limits_{i = 1}^n {{v_{i,t}}\bar \rho } } /nT = \sum\limits_{t = 1}^T {\sum\limits_{i = 1}^n {{v_{i,t}}} } /TL
\end{array} \right.
\end{equation}

\begin{figure}[ht!]
\centering
\includegraphics[scale=0.8]{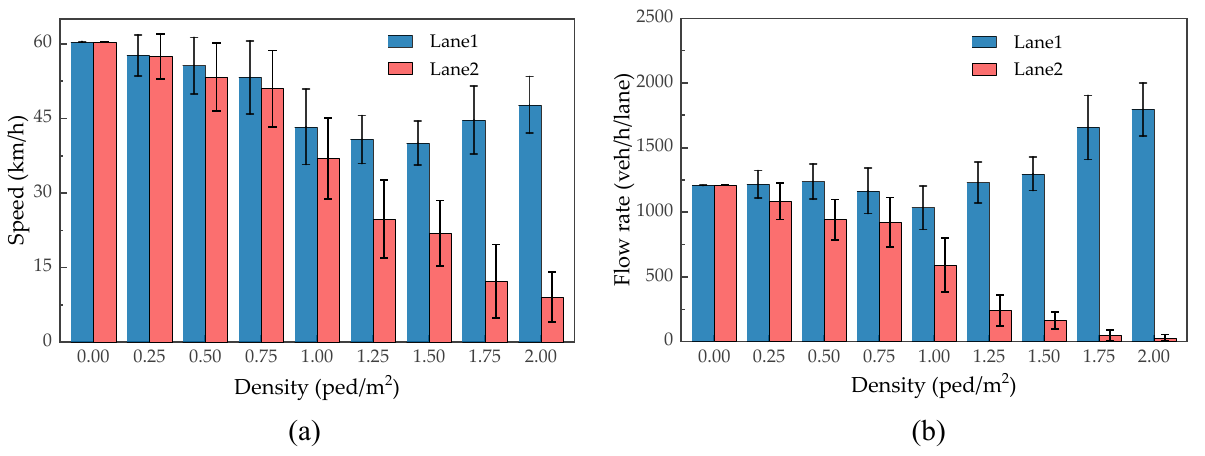}
\caption{Means (blue and red bars) with one standard deviation (black intervals) of speed and flow rate traffic indicators of the two lanes for different pedestrian densities (\({{\rho }_{veh}}\) =20 veh/km/lane).}
\label{fig18}
\end{figure}

\begin{figure}[ht!]
\centering
\includegraphics[scale=0.45]{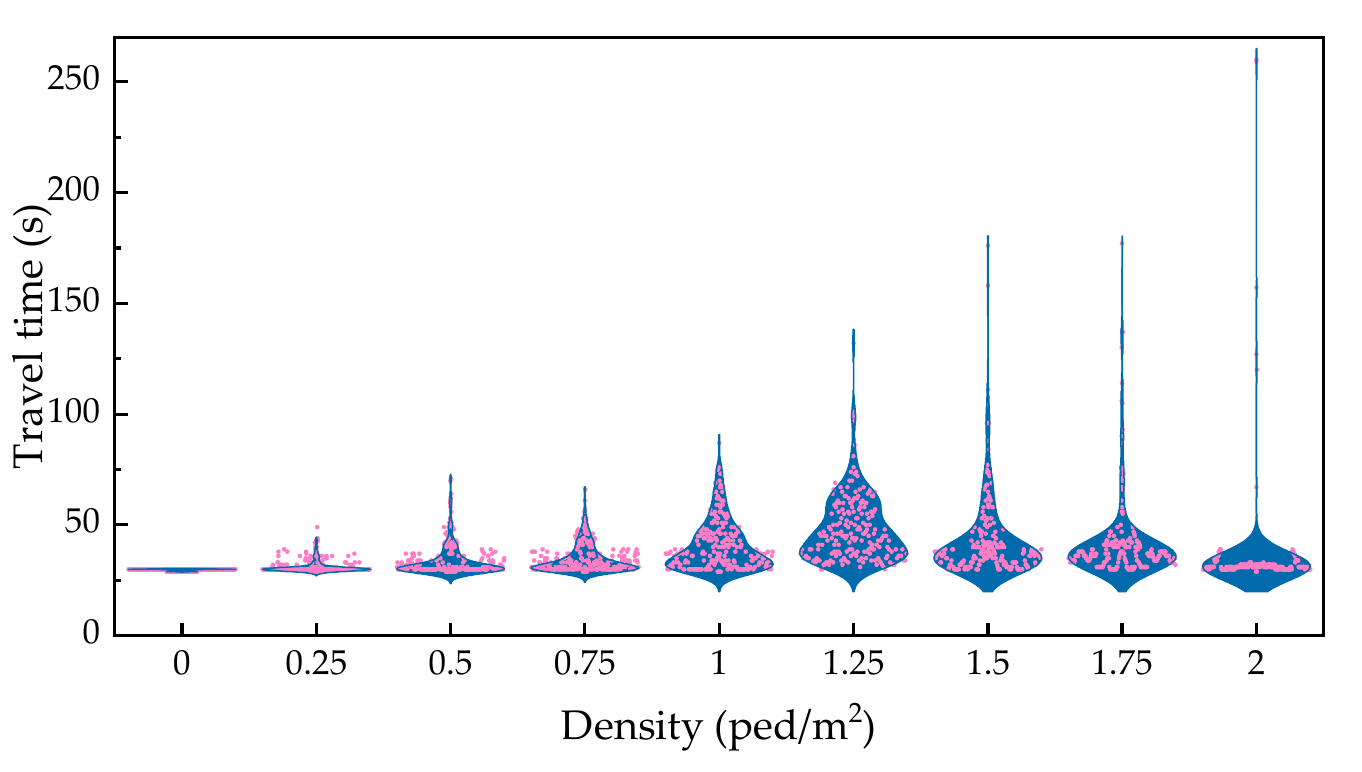}
\caption{Travel time distribution, the pink dots show the distribution of the data, and the contour lines in the blue area are the corresponding probability density fitting lines (\({{\rho }_{veh}}\) =20 veh/km/lane)}
\label{fig19}
\end{figure}

When the vehicles’ density is 20 veh/km/lane, by setting different pedestrian densities, the observed speed and flow rate changes in the two lanes are shown in Fig.\ref{fig18}. As we can see, with the increase in pedestrians’ density, the speed on both lanes has decreased significantly, and the downward trend in lane 2 is more significant. The speed drop in lane 2 is due to the intrusion of pedestrians, and the speed drop in lane 1 is caused by low-speed lane-changing vehicles. Fig.\ref{fig19}. shows the distribution of the travel time of the cycle of the vehicles in the simulation scenario of the periodic boundary. Corresponding to the speed change in Fig.\ref{fig18} (a), the mean travel time shows a rising trend and then falling. The shortest travel time is about 30 s, and the longest is more than 200 s.

\subsection{Saturated traffic (\({{\rho }_{veh}}=40 veh/km/lane\))} \label{subsection4.2}

\begin{figure}[ht!]
\centering
\includegraphics[scale=0.75]{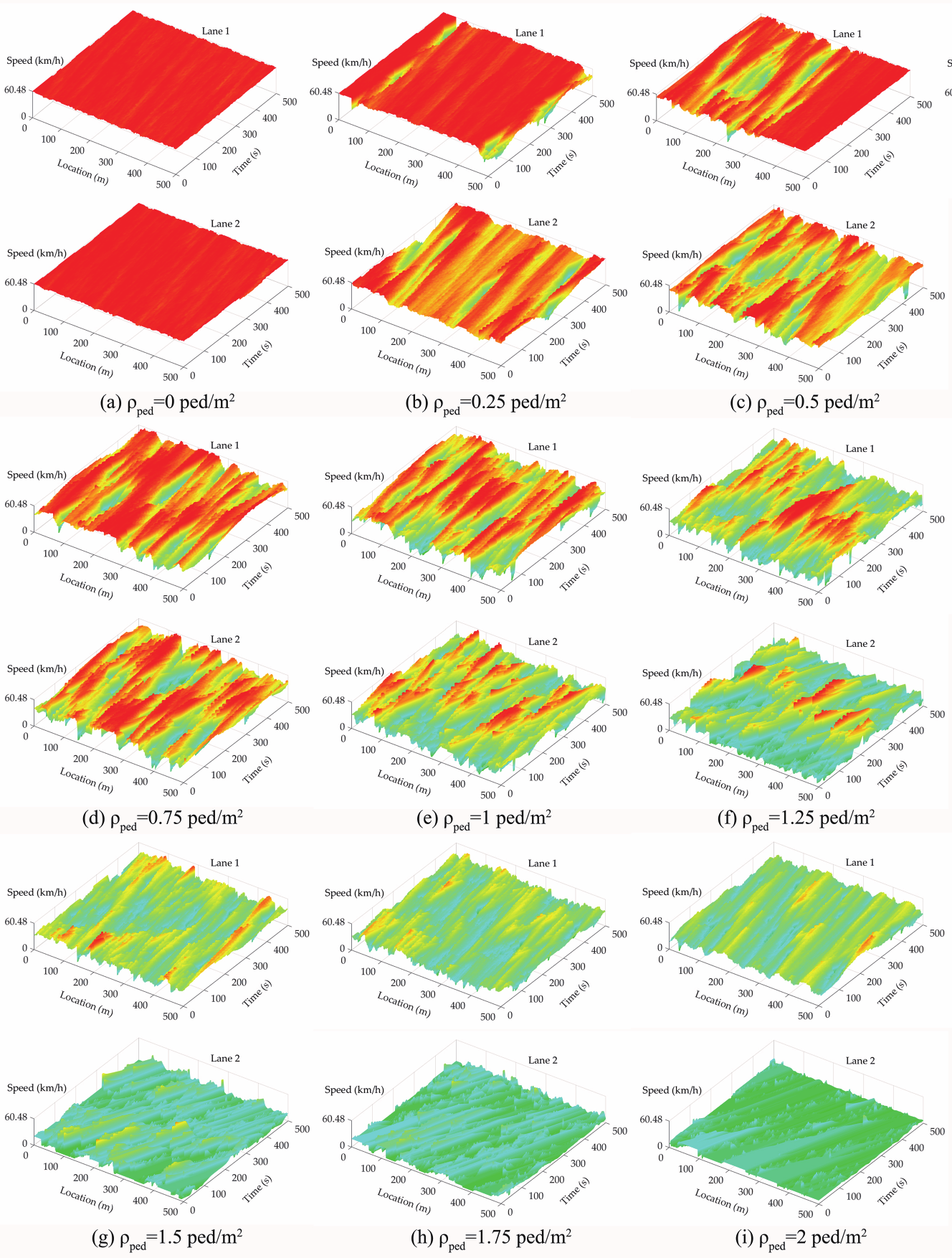}
\caption{Time-space evolution diagram of vehicles’ speed in 500 timesteps, \({{\rho }_{veh}}\)=40 veh/km/lane (We also provide a comprehensive two-dimensional time-space diagram version, please refer to the data availability section).}
\label{fig20}
\end{figure}

When the vehicles’ density increase to 40 veh/km/lane, the time-space diagram of vehicles’ speed under different initial pedestrians’ density states is shown in Fig.\ref{fig20}. Under this condition, the density of the vehicles reaches a saturated state, and the vehicles’ speed is close to free speed. However, with the increase in pedestrians’ density, intrusion behavior increases significantly, which seriously hinders the normal driving of vehicles. The vehicle movement exhibits an alternate state of low-speed following and stop-and-go. When the pedestrians’ density is higher than 1 ped/m\(^2\), the density of the two lanes exhibits a significantly uneven distribution, and many vehicles gather in lane 1.

\begin{figure}[ht!]
\centering
\includegraphics[scale=0.8]{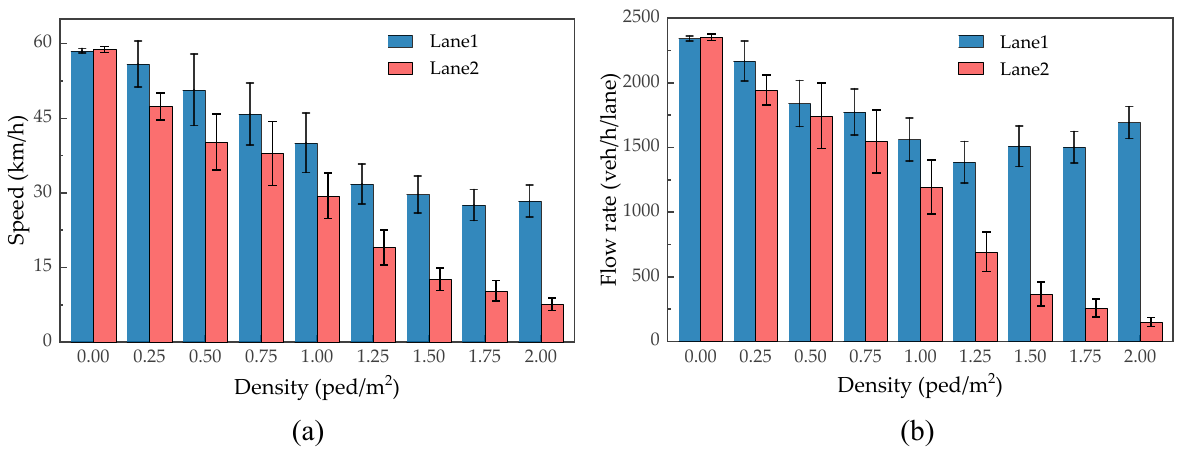}
\caption{Means (blue and red bars) with one standard deviation (black intervals) of speed and flow rate traffic indicators of the two lanes for different pedestrian densities (\({{\rho }_{veh}}\) =40 veh/km/lane)}
\label{fig21}
\end{figure}

\begin{figure}[ht!]
\centering
\includegraphics[scale=0.45]{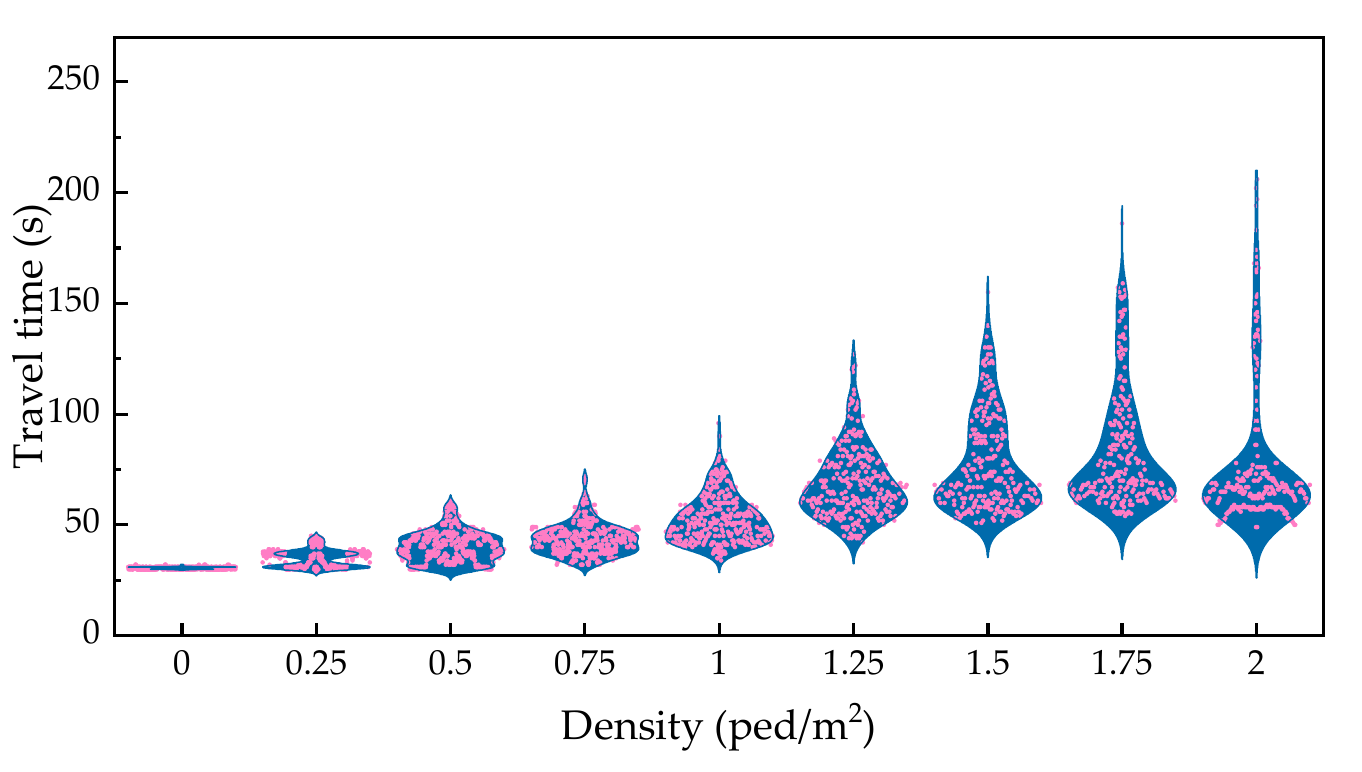}
\caption{Travel time distribution, The pink dots show the distribution of the data, and the contour lines in the blue area are the corresponding probability density fitting lines (\({{\rho }_{veh}}\) =40 veh/km/lane).}
\label{fig22}
\end{figure}

Fig.\ref{fig21} shows the two lanes' variation of vehicle mean speed and mean flow rate with one standard variation as a function of pedestrians’ density. When the pedestrians’ density is 0 ped/m\(^2\) (Fig.\ref{fig20} (a)), the vehicle runs in a saturated flow state, and the average speed is slightly lower than the maximum speed. As the density of pedestrians further increases, pedestrian intrusions increase and the speed difference between the two lanes becomes larger. Additionally, due to the uneven distribution of vehicles’ density on the two lanes, the difference in traffic flow states becomes more significant. From the perspective of changes in travel time (Fig.\ref{fig22}), when the pedestrians’ density is lower than 1 ped/m\(^2\), the travel time distribution is relatively stable. As the pedestrians’ density increases to 1 ped/m\(^2\), the travel time increases significantly, and the data distribution is more scattered, reflecting the uneven distribution of vehicles’ speeds in the two lanes.

\subsection{Oversaturated traffic (\({{\rho }_{veh}}\) =60 veh/km/lane)} \label{subsection4.3}

\begin{figure}[ht!]
\centering
\includegraphics[scale=0.75]{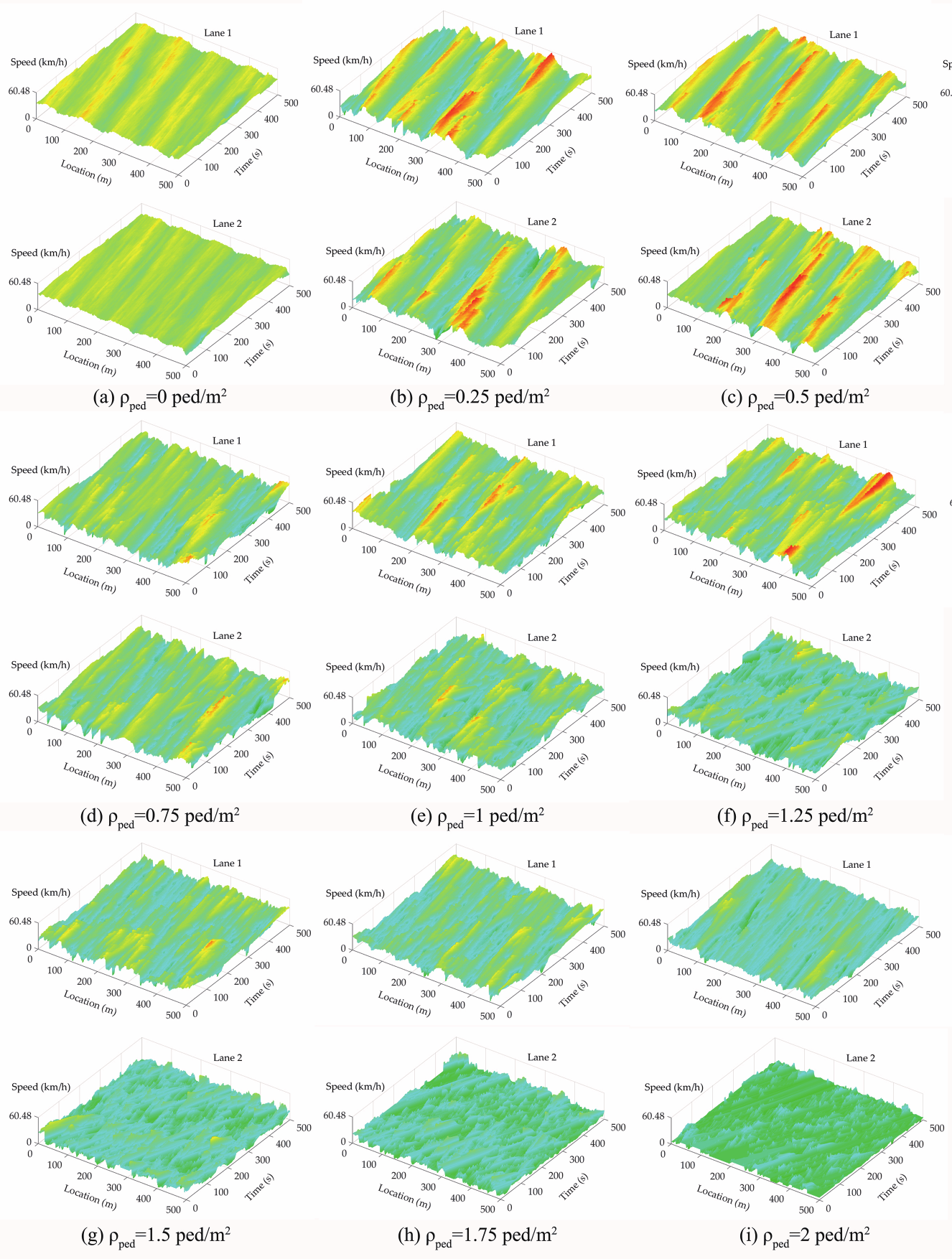}
\caption{Time-space evolution diagram of vehicles’ speed in 500 timesteps, \({{\rho }_{veh}}\)=60 veh/km/lane (We also provide a comprehensive two-dimensional time-space diagram version, please refer to the data availability section).}
\label{fig23}
\end{figure}

Comparison of different traffic flow states by setting equivalent increments of vehicles’ density, when the vehicles’ density is set to 60 veh/km/lane, the state of the traffic flow changes from free flow to synchronized flow, as shown in Fig.\ref{fig23}.

From Fig.\ref{fig23}, it can be found that along with the increase in pedestrians’ density, pedestrian intrusion and lane-changing behaviors have caused short-term local blockages in both lanes. However, due to the flexible lane-changing behavior of vehicles, the blockage state is difficult to develop into a large blockage group. The uneven traffic density distribution also appears in the simulation scenario when the pedestrians’ density further increases (Fig.\ref{fig23}(g)-(i)). The pedestrian intrusion behavior under the high vehicles’ density state is significantly reduced compared with the low vehicles’ density state.

\begin{figure}[ht!]
\centering
\includegraphics[scale=0.8]{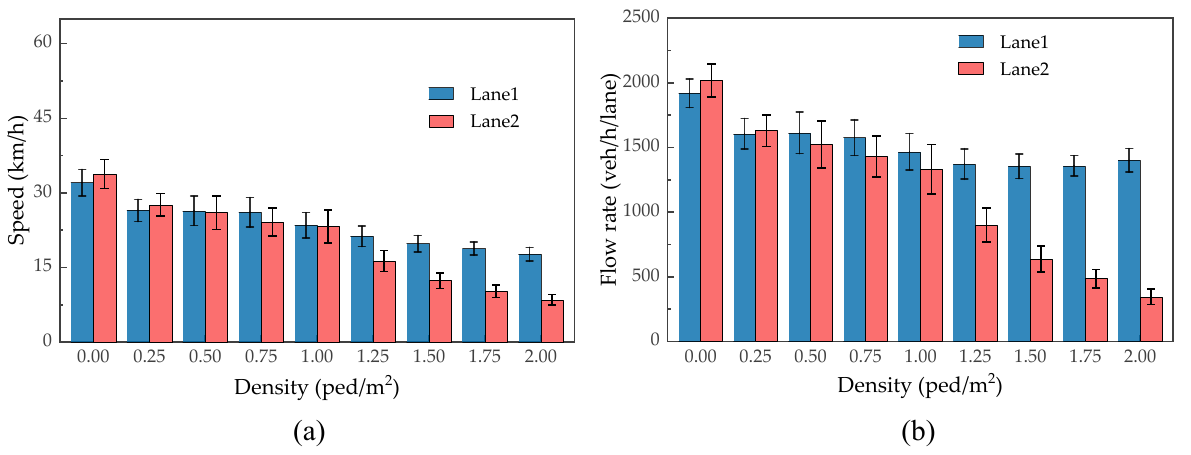}
\caption{Means (blue and red bars) with one standard deviation (black intervals) of speed and flow rate traffic indicators of the two lanes for different pedestrian densities (\({{\rho }_{veh}}\) =60 veh/km/lane).}
\label{fig24}
\end{figure}

\begin{figure}[ht!]
\centering
\includegraphics[scale=0.45]{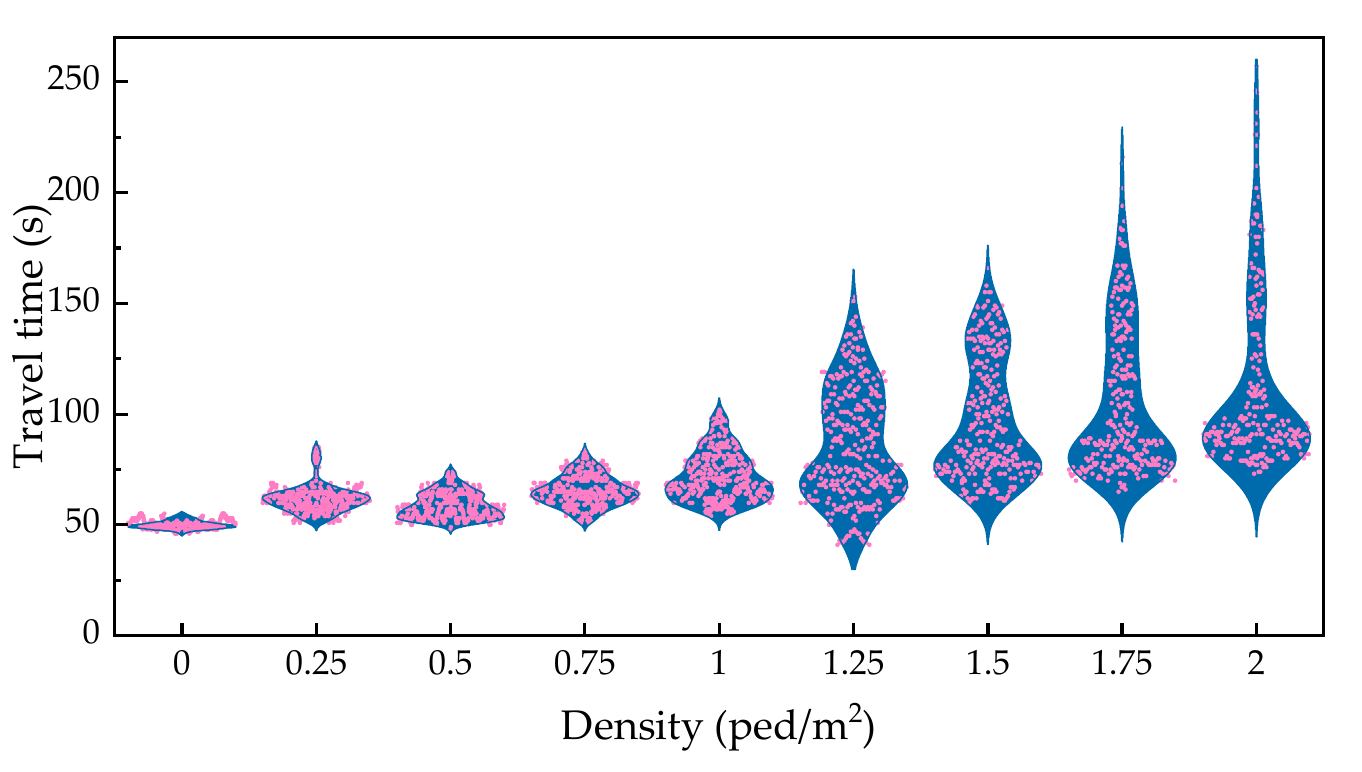}
\caption{Travel time distribution, The pink dots show the distribution of the data, and the contour lines in the blue area are the corresponding probability density fitting lines (\({{\rho }_{veh}}\) =60 veh/km/lane).}
\label{fig25}
\end{figure}

Observing the changes in the fundamental traffic parameters (Fig.\ref{fig24}), it can be noticed that when the pedestrians' density is between 0-1 ped/m\(^2\), the speed and flow rate in the two lanes are basically the same. When the pedestrians’ density further increases, the speed and flow rate on lane 2 decrease more significantly. From the relationship of travel time (Fig.\ref{fig25}), it can be seen that with the increase in pedestrians’ density, the travel time also witnessed a rise trend, and compared to the lower traffic density state (20, 40 veh/km/lane), the travel time is significantly increased.

\subsection{Congested traffic (\({{\rho }_{veh}}\) =80 veh/km/lane)} \label{subsection4.4}

\begin{figure}[ht!]
\centering
\includegraphics[scale=0.75]{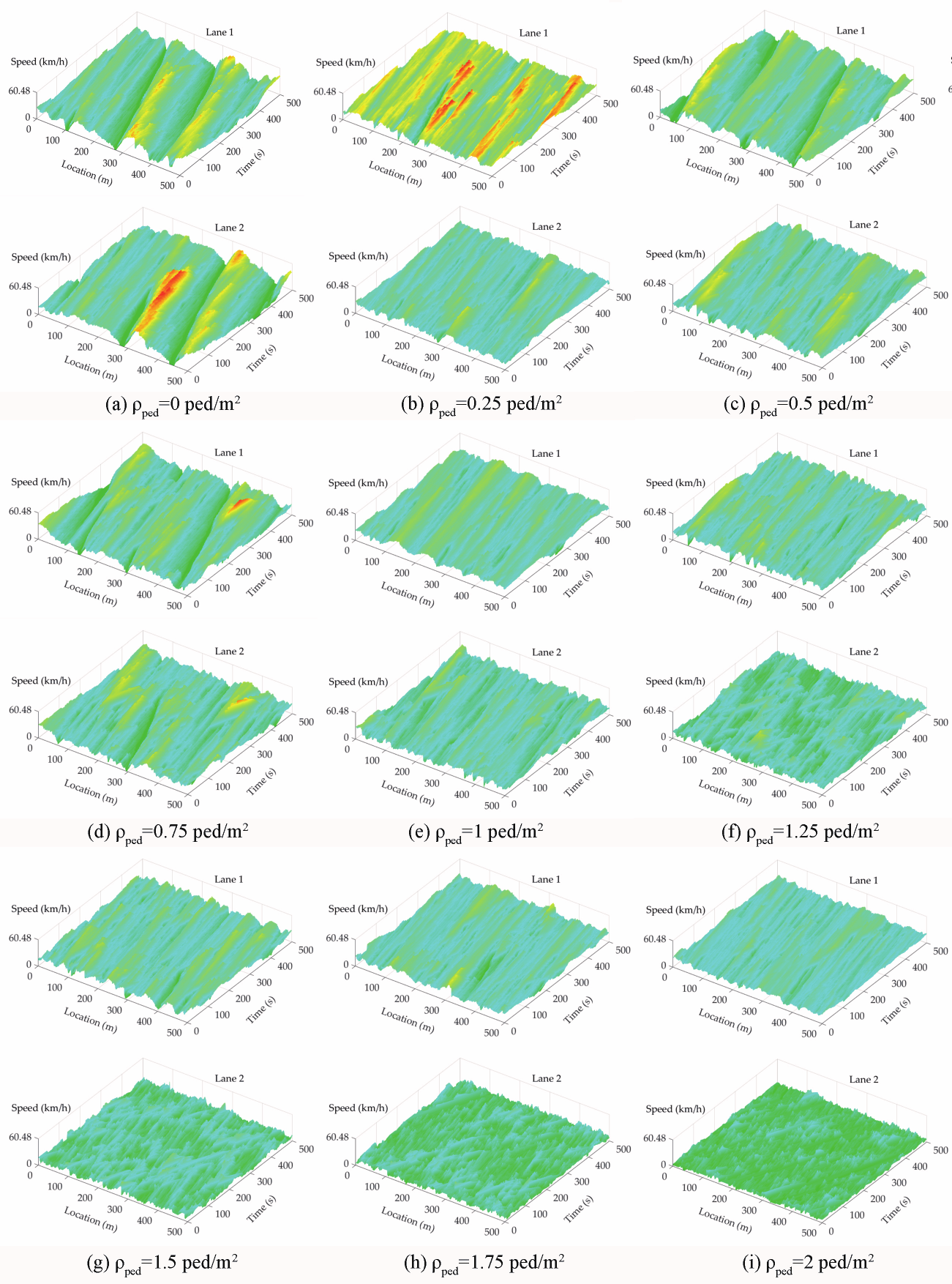}
\caption{Time-space evolution diagram of vehicles’ speed in 500 timesteps, \({{\rho }_{veh}}\) =80 veh/km/lane (We also provide a comprehensive two-dimensional time-space diagram version, please refer to the data availability section).}
\label{fig26}
\end{figure}

When the vehicles’ density is 80 veh/km/lane, as shown in Fig.\ref{fig26}, the traffic state in the simulation scenario is blocked. Due to the high traffic density, the vehicle follows the movement of the leader vehicle at a low speed.

It can be seen from Fig.\ref{fig26} (a)-(d) that when the pedestrians’ density is lower than 1 ped/m\(^2\), due to the mutual influence of the excessively high vehicles’ density and lane-changing vehicles on lane 1, a partial jam occurs moving upstream of the road.
When the pedestrians’ density is higher than 1 ped/m\(^2\), the traffic condition in lane 1 improves. This is because the excessive intrusion of pedestrians results in a differentiated density distribution on the two lanes. The high density of vehicles in lane 1 hinders lane-changing behavior, so it is less affected by lane-changing vehicles. In addition, because the vehicles' density in lane 2 is in the low level, low-speed vehicles in lane 1 can change to lane 2 in time, which also avoids congestion to a certain extent. From the perspective of pedestrian intrusion, as the density of vehicles increases, the frequency of pedestrian intrusion into lanes witnessed a downward trend.

\begin{figure}[ht!]
\centering
\includegraphics[scale=0.8]{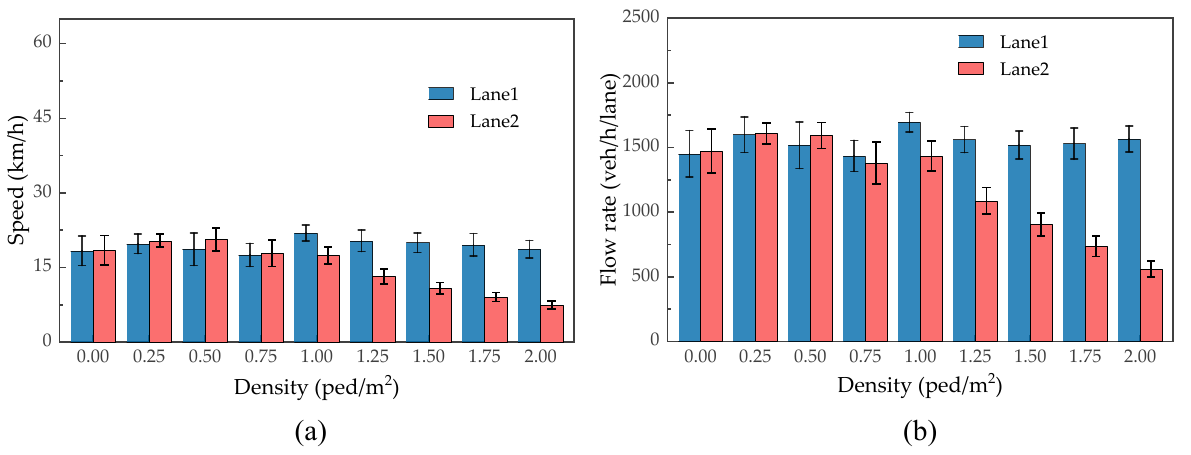}
\caption{Means (blue and red bars) with one standard deviation (black intervals) of speed and flow rate traffic indicators of the two lanes for different pedestrian densities (\({{\rho }_{veh}}\) =80 veh/km/lane).}
\label{fig27}
\end{figure}

\begin{figure}[ht!]
\centering
\includegraphics[scale=0.45]{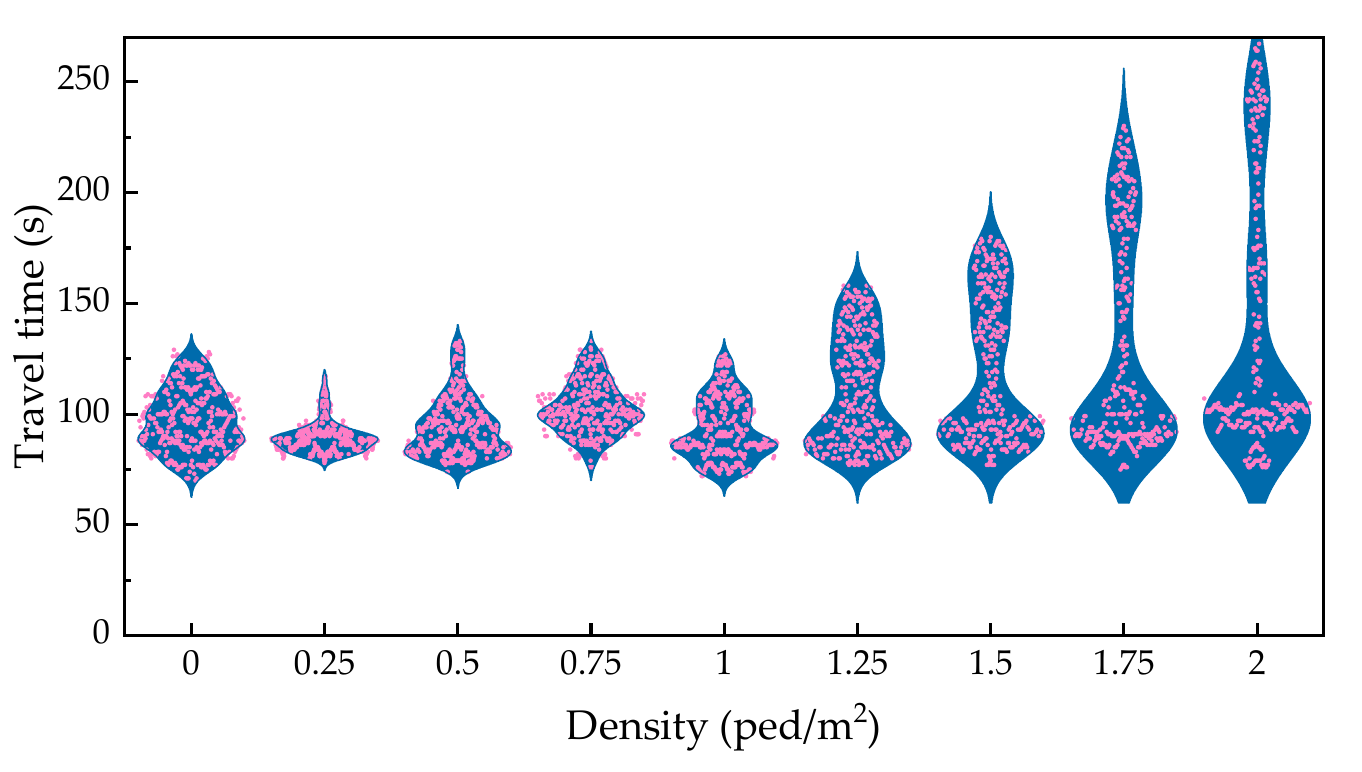}
\caption{Travel time distribution, The pink dots show the distribution of the data, and the contour lines in the blue area are the corresponding probability density fitting lines (\({{\rho }_{veh}}\) =80 veh/km/lane).}
\label{fig28}
\end{figure}

From the mean speed and mean flow rate statistics of the two lanes (Fig.\ref{fig27}), the speed change in lane 1 seems to remain stable, while the vehicles’ speed in lane 2 shows a continuous downward trend. The pedestrians’ density is 0.75 ped/m\(^2\) as a critical value. The speed relationship between the two lanes reverses when passing this point, and the traffic flow rate also shows a similar trend. Unlike the above situation, the distribution of travel time data (Fig.\ref{fig28}) is not concentrated in a specific interval, especially when the pedestrians’ density is higher than 1 ped/m\(^2\), the data distribution shows a trend of separation. The travel time data on each lane are concentrated in two different sections. This is because vehicles' lane-changing behavior becomes difficult under high-density conditions, so the difference in travel time between the two lanes is revealed.

\section{Fundamental diagram and phase diagram} \label{section5}

\subsection{Fundamental diagram} \label{subsection5.1}
This section analyses the fundamental and phase diagrams for different pedestrian and vehicle densities conditions. The simulation scenario and related parameters are the same as section \ref{section4}. The vehicles’ density was divided into 11 groups at intervals of 10 veh/km/lane (from 0 veh/km/lane to 100 veh/km/lane), and according to the pedestrians’ density, each group was divided into 9 groups at intervals of 0.25 ped/m\(^2\) (from 0 ped/m\(^2\) to 2 ped/m\(^2\)). Each set of data is averaged over 10 simulations to reduce the effect of the model error. A total of 980 independent simulations (Simulations of \({{\rho }_{veh}}=0, {{\rho }_{ped}}=0\) are excluded) were run and simulation data were obtained for each vehicles’ density and pedestrians’ density scenario. In each simulation, the first 1000 timesteps are discarded to omit the data of the transient times and the data are collected for the next 500 timesteps.

\begin{figure}[ht!]
\centering
\includegraphics[scale=1]{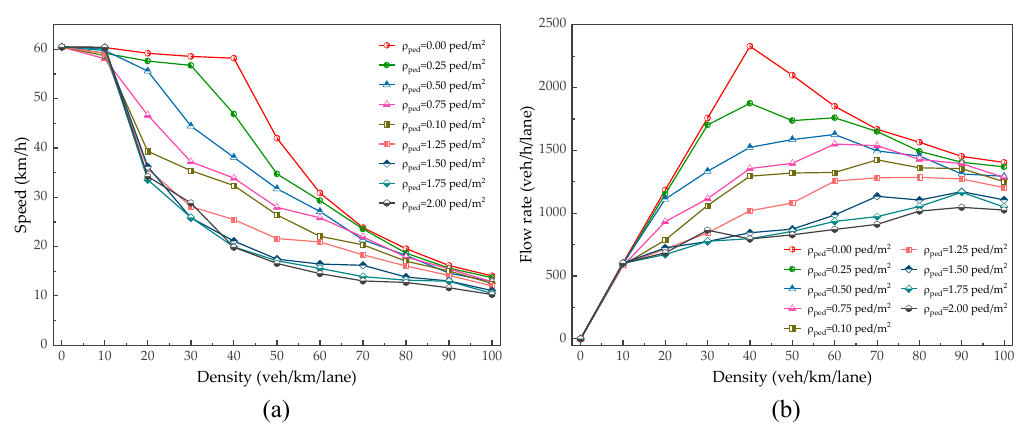}
\caption{Traffic fundamental diagram under different pedestrian densities, (a) Mean speed change as a function of the density of
vehicles for the considered pedestrian densities
, (b) Mean flow rate change as a function of the density
of vehicles for the considered pedestrian densities
.}
\label{fig29}
\end{figure}

Fig.\ref{fig29} shows the motorway's mean speed and mean flow rate change under different pedestrians’ density conditions on the sidewalk. When \({{\rho }_{ped}}=0\) ped/m2, the road traffic is in a homogeneous flow state. In this state, when the vehicles’ density is higher than 40 veh/km/lane, the vehicles’ speed begins to drop significantly, which means that the traffic flow state changes from free flow to synchronized flow. With the increase in pedestrians’ density, this state change has already occurred in the case of lower vehicles’ density. Of course, in a heterogeneous traffic flow, this has not changed from free flow to synchronized flow and traffic is oscillating due to the pedestrian intrusion. From the perspective of the mean flow rate change, when pedestrians’ density is below 1 ped/m\(^2\), the flow rate changes with 40 veh/km/lane as the critical point, i.e. mean flow rate starts to drop. When pedestrians’ density on the sidewalk is higher than 1 ped/m\(^2\), the curve of the mean flow rate diagram will rise and gradually stabilize with the increase of density, reflecting the completely different change between heterogeneous traffic and homogeneous traffic.

\begin{figure}[ht!]
\centering
\includegraphics[scale=0.9]{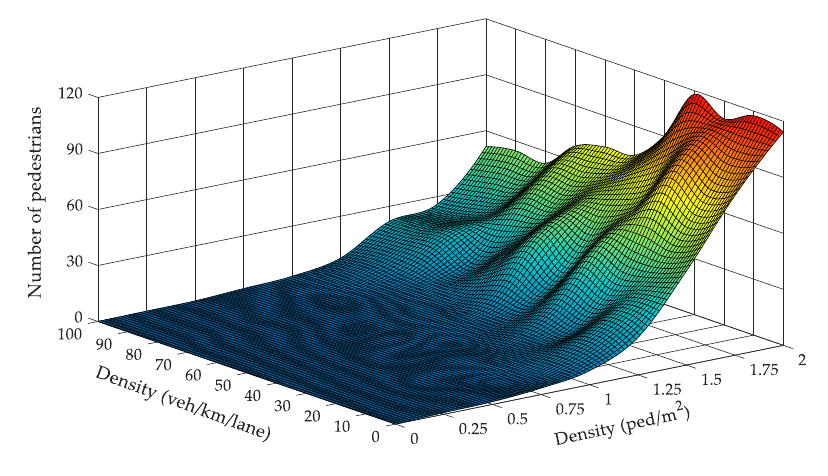}
\caption{Number of intruding pedestrians under different pedestrians’ and vehicles’ density.}
\label{fig30}
\end{figure}

Fig.\ref{fig30} shows the change in the number of intruders under different vehicles’ densities and pedestrians’ densities conditions in the last 500 timesteps. It can be seen from the figure that when the vehicles’ density are the same, the number of intruded pedestrians increases with the increase of the pedestrians’ density. When the pedestrians’ densities are the same, the number of intruded pedestrians decreases with the increase of the vehicles’ density. Taking \({{\rho }_{ped}}\)=1.0 ped/m\(^2\) as the critical point, when the pedestrians’ density is higher than this value, the number of intruding pedestrians begins to increase significantly, which means that pedestrians change from free flow to crowded flow at this density.

\subsection{Phase diagram} \label{subsection5.2}
\begin{figure}[ht!]
\centering
\includegraphics[scale=0.7]{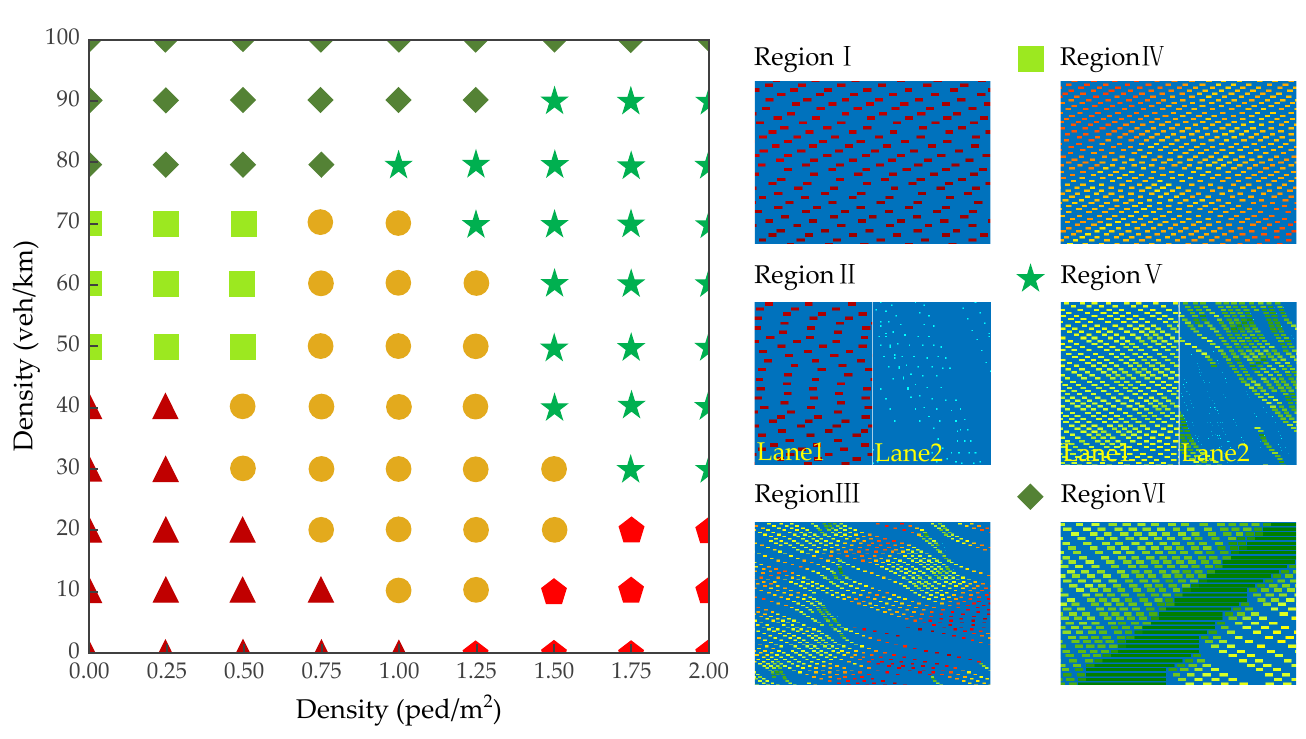}
\caption{Phase diagram of pedestrians’ densities and vehicles’ densities.}
\label{fig31}
\end{figure}

Analyzing the traffic state under different pedestrians’ density and vehicles’ density conditions can be roughly divided into six-phase regions, as shown in Fig.\ref{fig31}.

\textbf{Phase region \uppercase\expandafter{\romannumeral+1}, free-flow region} (pedestrian flow state: free flow, traffic flow state: free flow): in this region, there are few pedestrian intrusions phenomenon, and vehicles can drive in a free flow state.

\textbf{Phase region \uppercase\expandafter{\romannumeral+2}, uneven free-flow region} (pedestrian flow state: congestion flow, traffic flow state: free flow): excessive intrusion by pedestrians has caused extreme density differences between lanes in this region. Vehicles in lane 1 drive in free flow and gather there, while lane 2 is mainly occupied by pedestrians, with fewer vehicles.

\textbf{Phase region \uppercase\expandafter{\romannumeral+3}, oscillating flow region} (pedestrian flow state: congested flow, traffic flow state: free flow to synchronized flow): in this region, many pedestrian intrusions on the road cause frequent traffic conflicts. There are numerous disturbances in the traffic flow, and lane-changing behaviors are frequent.

\textbf{Phase region \uppercase\expandafter{\romannumeral+4}, synchronized flow region} (pedestrian flow state: free flow, traffic flow state: synchronized flow): in this region, there are fewer pedestrian intrusions, and the traffic is in a stable synchronized flow state.

\textbf{Phase region \uppercase\expandafter{\romannumeral+5}, homogeneous congested region} (pedestrian flow state: congested flow, traffic flow state: synchronized flow): in this region, traffic is very congested, and vehicles travel in synchronized flow at low speed.

\textbf{Phase region \uppercase\expandafter{\romannumeral+6}, moving jam region} (pedestrian flow state: free flow to congested flow, traffic flow state: moving jam): in this region, the traffic flow is in a moving jam state, and because vehicle congestion hinders pedestrians’ intrusion behavior, there are fewer pedestrian intrusions.

\section{Traffic conflict analysis} \label{section6}

The intrusion of pedestrians significantly impedes traffic flow and frequently results in numerous vehicular conflicts. On urban streets, incidents involving pedestrian-vehicle interactions constitute a substantial proportion of road accidents. Typically, these collisions transpire at junctions, pedestrian crossings, or on roadways experiencing considerable pedestrian incursions. This section provides a quantitative analysis of pedestrian-vehicle conflicts within mixed traffic environments involving both pedestrians and vehicles.

Vehicles’ conflict analysis is the focus of traffic safety research. The time headway distribution, TTC, deceleration response time, maximum deceleration rate, and other statistics are used to measure the deceleration state of the vehicle. In the traffic conflict prediction model of \citet{katrakazas2021prediction}, the TTC and Deceleration Rate to Avoid Crash (DRAC) metrics are utilized to estimate conflicts’ frequency, \citet{bosetti2014human} pointed out that based on a large amount of observational, when longitudinal decelerations is greater than 3 m/s\(^2\), true braking is thus needed. Deceleration rates proposed/observed by most researchers are less or equal to the deceleration rate proposed by \citet{manual2010american}, with a comfortable deceleration rate of 3.4 m/s\(^2\). So, we define a traffic conflict as occurring when the vehicle's deceleration exceeds 3.6 m/s\(^2\) (9\(\cdot\delta a\)).

In this part, we analyze the traffic conflict frequency under different pedestrians’ and vehicles’ densities. The simulation scenarios are consistent with section \ref{section4}. The relevant parameters of the model are shown in Tab.\ref{table2}, Tab.\ref{table3}, and Tab.\ref{table4}. The vehicles’ densities was divided into 5 groups at intervals of 20 veh/km/h ((from 20 veh/km/lane to 100 veh/km/lane)), and each group of pedestrians was divided into 9 groups at intervals of 0.25 ped/m\(^2\) (from 0 ped/m\(^2\) to 2 ped/m\(^2\)). Each set of simulations was repeated 10 times to reduce the impact of model errors. A total of 440 independent simulations (Simulations of \({{\rho }_{veh}}=0, {{\rho }_{ped}}=0\) are excluded) were run and obtained the traffic conflicts’ frequency data for each vehicles’ density and pedestrians’ density scenario. In each simulation, the first 1000 timesteps are discarded to omit the data of the transient times and the data are collected for the next 500 timesteps.

\subsection{Speed limit} 
\label{subsection6.1}

\begin{figure}[ht!]
\centering
\includegraphics[scale=1]{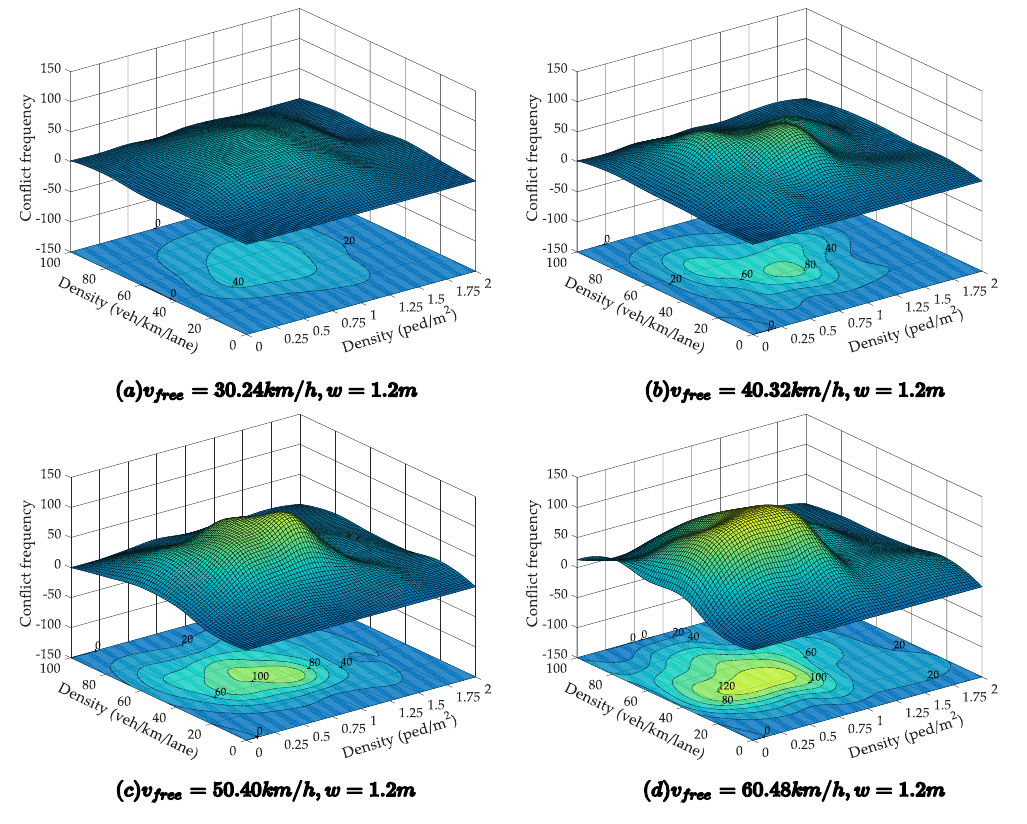}
\caption{Conflict frequency distribution under different speed limits (\({{v}_{free}}\)).}
\label{fig32}
\end{figure}

In urban traffic, road speed limits are crucial for reducing the frequency and severity of accidents.  In our simulation, we adjusted the free-flow speeds to emulate various speed limit conditions and examined the distribution of conflict frequency across different densities of vehicles and pedestrians, as illustrated in Fig.\ref{fig32}. Additionally, the parameters of the car-following model under different speed limit conditions are detailed in Tab.\ref{table2}.

As seen from Fig.\ref{fig32}, under different speed limit conditions, the distribution of conflict frequency shows a significant difference. Higher driving speed, on the one hand, improves traffic efficiency, on the other hand, it will also cause more traffic conflicts. When the speed limit is 60.48 km/h, the peak conflict frequency is close to 140, which is three times higher than when the speed limit is 30.24 km/h (the corresponding conflicts’ frequency peak value is about 40). Within a specific density interval (vehicles’ density between 20-80 veh/km/lane, pedestrian’s density between 0.25-1.25 ped/m\(^2\)), the conflicts’ frequency is much higher, and the change of conflict frequency shows strong fluctuations.

\subsection{Sidewalk width} \label{subsection6.2}

\begin{figure}[ht!]
\centering
\includegraphics[scale=1]{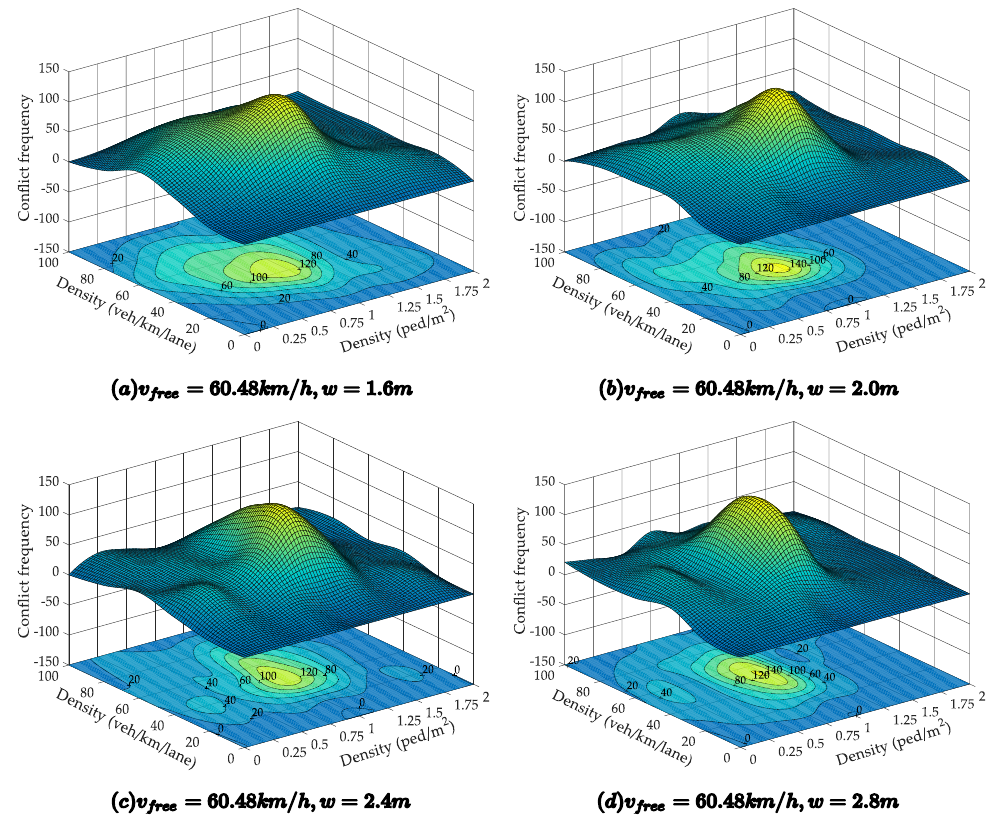}
\caption{Conflict frequency distribution under different sidewalk widths (\(w\)).}
\label{fig33}
\end{figure}

Furthermore, we analyzed the impact of sidewalk width on conflict frequency. Sidewalk widths were set at 1.6 m, 2.0 m, 2.4 m, and 2.8 m. The distribution of conflict frequency under various sidewalk widths is illustrated in Fig.\ref{fig33}, while the conflict frequency distribution for a 1.2 m sidewalk width is presented in Fig.\ref{fig32}(d).

From the data, it is evident that increasing the sidewalk width does not significantly alter the peak values of conflict frequency, which remain within the range of 130-140. Notably, in certain pedestrian density intervals (0-0.5 ped/m\(^2\) and 1.25-2 ped/m\(^2\)), there is a marked decline in conflict frequency as the sidewalk width increases. Another observation from the figure is the clustering of contour lines, indicating a centralization in data distribution as the sidewalk width increases. In specific density areas, the conflict frequency markedly increases, whereas it remains low outside these areas.

Wider sidewalks accommodate more pedestrians, but the frequency of collisions does not escalate as a result. Contrarily, within certain density ranges, the collision frequency actually decreases as the sidewalk width expands. These findings suggest that wider sidewalks can effectively reduce conflicts between pedestrians and vehicles.

\section{Conclusions} \label{section7}

This paper introduces a cellular automata model designed for mixed pedestrian-vehicle traffic. Both pedestrians and vehicles are unified in terms of spatial dimensions and simulation rules, ensuring compatibility between submodels. The simulations of this model provide accurate representations at both macroscopic and microscopic levels.

By using this model to simulate pedestrian-vehicle mixed traffic flows, this paper initially examines the speed and flow rate statistics in two lanes under varying vehicle densities. The results indicate that vehicles are more likely to change lanes to avoid the low-speed driving conditions caused by pedestrian intrusion in states of low vehicle density. In contrast, at high vehicle densities, it becomes difficult for vehicles in lane 2 to switch to lane 1 because of pedestrian intrusion. Consequently, with an increase in pedestrian density, long-range jams and stop-and-go waves emerge in lane 2. Subsequent, the analysis of the fundamental diagram reveals distinct flow rate variations in heterogeneous traffic compared to homogeneous traffic, attributed to pedestrian intrusions that force vehicles to decelerate before reaching saturation density. Additionally, phase diagram analysis of heterogeneous traffic identifies six distinct regions based on traffic conditions: free-flow state, unevenly distributed free-flow state, intensive traffic conflict state, synchronized flow state, unevenly distributed synchronized flow state, and moving jam state.

Through simulation, this paper also explores the frequency of vehicle conflicts under various speed limits and sidewalk widths. Simulation results demonstrate that stricter speed limits significantly reduce the frequency of collisions, for instance, at a speed limit of 30.24 km/h, the collision frequency is one-third of that at 60.48 km/h. With lower pedestrian densities, wider sidewalks contribute to reduced road conflicts, peaking when pedestrian density reaches 1 ped/m\(^2\).

However, several issues warrant further attention in future work. To simplify the modeling, the behavior of vehicles maneuvering to avoid pedestrians was not considered, and the conflict rules between pedestrians and vehicles were also simplified. Future modifications to the model aim to enhance its accuracy and apply it to more complex mixed traffic scenarios.

\centerline{}
\section*{Data Availability}\label{data}
The data and codes can be found here: {\url{https://drive.google.com/drive/folders/1NYVnRp0z8VPuskfezMr51gB-sraOf6Iq?usp=drive_link}} (Google Drive).

\centerline{}
\section*{Acknowledgments}
This work was supported by the National Natural Science Foundation of China (Grant No. 52072286, 72074149), the Opening Fund of State Key Laboratory of Fire Science (Grant No. HZ2021-KF11) and the Fundamental Research Funds for the Central Universities (Grant No. 2020VI002).

%\bibliography{sample631}{}
\bibliographystyle{aasjournal}

\end{document}